\documentclass[12pt]{article}
\usepackage{amssymb}
\usepackage{amsmath}
\usepackage{graphicx}
\usepackage{graphics}

\newcommand{\ie}{i.e.}
\newcommand{\nn}{n.n.}
\newcommand{\nnn}{n.n.n.}

\title{On the nature of striped phases:
Striped phases as a stage of ``melting'' of 2D crystals}

\author{Volodymyr Derzhko$^1$, Janusz J{\c{e}}drzejewski$^1$\thanks{Corresponding author: J. J{\c{e}}drzejewski, phone:
+48 71 3759415, fax: +48 71 3214454, e-mail: jjed@ift.uni.wroc.pl} and Taras Krokhmalskii$^2$\\
$^1$ Institute of Theoretical Physics, University of Wroc{\l}aw,\\
pl. Maksa Borna 9, 50--204 Wroc{\l}aw,
Poland \\
$^2$ Institute for Condensed Matter Physics,\\
1 Svientsitskii Str.,
Lviv-11, 79011, Ukraine}

\begin{document}

\maketitle

\begin{abstract}
We discuss striped phases as a state of matter intermediate between
two extreme states: a crystalline state and a segregated state. We argue that
this state is very sensitive to weak interactions, compared to those
stabilizing a crystalline state, and to anisotropies.
Moreover, under suitable conditions a 2D system in a striped phase decouples
into (quasi) 1D chains. These observations are based on results of our studies
of an extension of a microscopic quantum model of crystallization, proposed originally
by Kennedy and Lieb.
\end{abstract}

\section{Introduction}

Striped phases, specific phases having quasi-one-dimen\-sio\-nal
structure, are ubiquitous. They have been observed in numerous
experiments carried out in a variety of systems, among which there
are physisorbed monolayers on metallic surfaces \cite{kern} or
ultrathin magnetic garnet films \cite{allenspach,seul}. They have
also been found in theoretical analyses, carried out by analytic or
numerical methods, of various quantum and classical model systems.
In the class of quantum models, one can find results pertaining to
the Hubbard model \cite{poilblanc,machida,oles}, $t$--$J$ model
\cite{white1,white2}, $XY$ model \cite{sandvik}, spinless fermion
model \cite{zhang1,zhang2}, spinless Falicov--Kimball model
\cite{lemanski1,DJ3}. It has been argued in a number of works that
some essential physics of systems displaying striped phases can be
grasped by a kind of coarse-grained descriptions that result in
classical lattice-gas models (Ising-like models or $O(n)$ models)
with competing two-body interactions. Typically, in these models
short-range attractive (ferromagnetic) interactions compete with
long-range repulsive ones \cite{Sasaki1,Low,BoothMWB1,ArlettWMB1,StoychevaS,ValdezS1}.
Interestingly, despite the difficulty of determining ground states of such systems,
in recent years a few rigorous results have been published
\cite{nussinov,giuliani123}.

Perhaps the most notable striped phases are those observed in doped
layered perovskites, which under suitable conditions become
high-temperature superconductors \cite{tranquada1,tranquada2}. The
nature of striped phases in these materials, in particular their
competition with a superconducting state, is still vigorously
debated.

The purpose of this paper is threefold. Firstly, we argue that the
problem of the formation of striped phases falls naturally into the
context of an extended crystallization problem, that is not only the
existence of crystalline phases is addressed but also the process of
their deterioration into a segregated state (described below) as a
result of weakening of the forces stabilizing a crystalline state. A
crystalline phase and a segregated one might be thought of as two
extreme states of matter. Presumably, striped phases appear in this
process as the last but one stage of a sequence of phase
transformations that starts with a crystalline state and ends with a
segregated one.

Secondly, we point out that the stability of striped phases might be
rather fragile. Very weak interactions, in the hierarchy of the
interactions present in the system under consideration,
can change the character of the striped pattern or destabilize completely striped
phases. Moreover, even a weak anisotropy of the system chooses the direction
of stripes in stable striped phases.

Thirdly, the appearance of striped phases might signal a dramatic destruction
of correlations in direction of stripes. An anisotropy of the system
favors this effect. Specifically, we show that when our system is in an
axial-stripe phase, zero-temperature electron correlations
(given by ground-state one-body reduced density matrix) along the stripes vanish.
We make an attempt to estimate the range of temperatures
in which this property holds approximately.
Consequently, before a 2D crystal deteriorates completely
(to a state of segregation), it transforms effectively into a quasi
one-dimensional structure, destroying some of existing 2D long-range orders.

Our main conclusions and suggestions are based on a rigorous
analysis of an extension of a microscopic quantum model of
crystallization, which was proposed twenty two years ago by Kennedy
and Lieb \cite{kennedy1}. Some additional observations are based on
numerical calculations. In brief, the Kennedy and Lieb model of
crystallization is a two-component fermionic system on a lattice
that consists of heavy immobile ions and light hopping electrons,
and the sole interaction in this system is an on-site electron-ion
interaction representing a screened Coulomb interaction.

The paper is organized as follows. In the next section we introduce
our model of crystallization. Then, in Section 3, we present some of the ground-state
phase diagrams of our system, discuss possible scenarios of deterioration of
a checkerboard-like 2D crystal, emphasizing the role of striped phases and
the effect of an anisotropy of electron hopping.
After that, in Section 4, we focus our attention on axial-stripe phases,
specifically on electron correlations in such phases.
Finally, in Section 5, we summarize our results.
Spectral properties of axial-stripe phases are presented in Appendix.

\section{The model of crystallization}

The model of crystallization proposed by Kennedy and Lieb
\cite{kennedy1} is composed of two fermionic subsystems: light
hopping electrons and heavy immobile ions. The electrons are
represented by spinless fermions (spin does not play any role in our
considerations), described by creation and annihilation operators of
an electron at a state localized at site $x$ of the underlying
lattice, $c^{+}_{x}$, $c_{x}$, respectively, satisfying the
canonical anticommutation relations. The ions are described by
collections of pseudo-spins $\left\{ s_{x} \right\}_{x \in
\Lambda}$, called the {\em ion configurations}; $s_{x}=1$ if the
site $x$ is occupied by an ion and $s_{x}=-1$ if it is empty. The
pseudo-spins commute with the creation and annihilation operators of
electrons.

There is neither a direct interaction between mobile electrons nor
between immobile ions. The electrons energy is due to hopping
(typically a nearest-neighbor (\nn{}) hopping is assumed) with $t$
being the \nn{} hopping intensity (without any loss of generality we
can fix the sign of $t$, $t>0$), and due to a screened Coulomb
interaction with the immobile ions, whose strength is controlled by
the coupling constant $U > 0$ (concerning the sign see a comment in
the text below). The Hamiltonian of the system reads:

\begin{eqnarray}
H_{FK}  = t \hspace{-1mm} \sum\limits_{\langle x,y \rangle_{1}} \hspace{-1mm}
\left(
c^{+}_{x}c_{y}+c^{+}_{y}c_{x} \right) + U\sum\limits_{x}\left(
c^{+}_{x}c_{x} - \frac{1}{2} \right) s_{x} ,
\label{HFK}
\end{eqnarray}
where $\langle x,y \rangle_{1}$ means that the sites $x,y$ constitute a
pair of \nn{} sites.
In the sequel, we shall limit our considerations only to the case
in which the underlying lattice is a square lattice.

It is worth to emphasize here that although the ions are not moving due to
the dynamics of the system, their  configurations are not frozen or random.
On calculating the canonical partition function, one takes a trace over electronic
degrees of freedom and sums over all configurations of the ions.
It is this sampling of possible configurations that produces correlations between
ions, which in turn may lead to a crystalline arrangement of the ions.

The Hamiltonian $H_{FK}$ is widely known as the Hamiltonian of the
spinless Falicov--Kimball model, a simplified version of the
Hamiltonian put forward  for describing electronic subsystems of some solids
in \cite{FK}.
To the best of our knowledge, the spinless-fermion Falicov--Kimball model is
the unique system of interacting fermions, for which the existence of long-range orders
(periodic phases)  and phase separations have been proved.
The majority of rigorous results refers to the ground states of the model
and holds only in the strong-coupling regime, \ie{} for sufficiently small $|t/U|$,
and for particle densities satisfying specific conditions (some more details and comments
can be found below).
Interestingly, the first proofs of the existence of a periodic phase,
which in 2D is the so called checkerboard phase with the electron and ion densities equal $1/2$,
hold for any $U$: at zero temperature \cite{brandt,kennedy1} and for sufficiently low
temperatures \cite{kennedy1}.
It is the checkerboard phase that in our considerations
is identified with an initial crystal, which is given a possibility to deteriorate.
A review of results pertaining to ground-state or low-temperature phase
diagrams and an extensive list of relevant references can be found in \cite{GM,JL}).

According to the state of art, a quite general analysis of
ground-state phase diagrams, that made possible proving the
existence of numerous periodic phases and mixtures (phase separated
states) of such phases, is feasible only in the mentioned above
strong-coupling regime, and when the densities of the ions and
electrons, $\rho_i$ and $\rho_e$, respectively, satisfy a specific
condition. Either it is the condition of neutrality, $\rho_i =
\rho_e$, if the electron-ion interaction is attractive, or it is the
condition of half-filling, $\rho_i + \rho_e =1$, if that interaction
is repulsive. The two cases are related by a unitary transformation:
a hole-particle transformation (for definiteness, we assume in our
considerations that the electron-ion interaction is repulsive and
the system is half-filled). In the specified above regime, it is
possible to derive explicitly, in the form of a convergent power
series with respect to the small parameter $|t/U|$, an effective
interaction between the ions \cite{kennedy2,MM-S,GMMU,DFF}. The
components of this expansion constitute many-body finite-range
lattice-gas interactions, where the number of interacting bodies and
the range of interaction grow without bound with the order of those
components. Then, it is possible to construct rigorously the
ground-state phase diagram of the effective interaction truncated at
certain order, and after that derive information on the ground-state
phase diagram of the complete quantum system under consideration
\cite{kennedy2,MM-S,GMMU,DFF}. The ground-state results obtained in
the described way can be extended to sufficiently low temperatures
\cite{DFF}.

The emerging phase diagram is very reach; besides a few periodic
phases \cite{GMMU,kennedy3,haller1,haller2} and various
phase-separated states \cite{kennedy3,haller1}, that can be
determined rigorously, it contains most probably infinitely many
periodic phases and  mixtures of periodic phases (see
\cite{GUJ,GJL,WL,fark} for the results obtained by means of the
method of restricted phase diagrams). However at half-filling and
large $U$, it does not contain any segregated phases, that is
thermodynamic mixtures of the completely filled with ions phase
($\rho_i =1$) and the ion void phase ($\rho_i =0$) with suitable
electron densities. This is in agreement with rigorous results of
\cite{Lemberger,FLU1,FLU2}, which state that a segregated phase is
stable for any total density away from half-filling if $U$ is large
enough. Approximate results of \cite{GUJ,GJL,WL} suggest that
whatever $U>0$, there is no segregated phases at half-filling.

We insist, however, that a crystalline state, like the half-filled
checkerboard phase, should have a possibility to deteriorate into
the half-filled segregated phase, considered as a final stage of a
deterioration of a crystal, for any $U$. The standard
Falicov-Kimball model cannot encompass such a ``process'': there is
neither the stable half-filled segregated phase nor a parameter to
drive such a process. The reason for which half-filled segregated
phases are missing in the standard spinless Falicov--Kimball model,
in the strong-coupling regime, might be attributed to the fact that
the leading interaction term (whose order is $O(|t/U|^2)$) of the
effective interaction expansion is a {\nn} repulsive lattice gas (or
in terms of pseudo-spins --- an Ising antiferromagnet),
irrespectively of the sign of $U$. In order to stabilize the
half-filled segregated phase and to provide the standard model with
a suitable control parameter, we extend it by adding a small (second
order as compared to the electron-ion interaction) short-range
attractive interaction between the ions. For simplicity, we choose
its range to be limited to the distance between
next-nearest-neighbor ({\nnn}) sites. The physical source of this
interaction might be identified as  van der Waals forces, which
despite their weakness are known to play an important role in
stability of crystals. Therefore, we extend the standard spinless
Falicov--Kimball model by adding to the Hamiltonian $H_{FK}$ the
term $H_{vdW}$,

\begin{eqnarray}
H_{vdW}=\frac{W}{8} \sum\limits_{\langle x,y \rangle_{1}} s_{x}s_{y}
-\frac{\tilde{\varepsilon}}{16} \sum\limits_{\langle x,y
\rangle_{2}} s_{x}s_{y},
\label{VdW}
\end{eqnarray}
with
\begin{eqnarray}
\label{new-var} W = -2t^{2}+ t^{4}\omega ,
\hspace{5mm}
\tilde{\varepsilon} = t^{4} \varepsilon ,
\end{eqnarray}
where $\langle x,y \rangle_{2}$ stands for a pair of \nnn{} sites.
The parameters $\omega$ and $\varepsilon$, which vary the strength
of \nn{} and  \nnn{} couplings in $H_{vdW}$, respectively, are the
control parameters of the phase diagrams discussed in the sequel.
The specific form of $W$, given in (\ref{new-var}), and the
coefficients that stand by the parameters $W$ and
$\tilde{\varepsilon}$ in (\ref{VdW}) stem from our analysis of the
ground-state phase diagram of the 2nd order effective interaction
(see \cite{DJ3} for details). In those diagrams,
$\omega=\varepsilon=0$ is the point of coexistence of the
checkerboard phase and the half-filled segregated phase. Despite the
weakness of the \nnn{} term in $H_{vdW}$, it will become clear in
the discussion of phase diagrams that follows, that extending the
range of $H_{vdW}$ beyond \nn{} enables us to make interesting
observations pertaining to a transition from a crystalline state to
a segregated one.

To take into account important effects of anisotropy of electron hopping,
we differentiate between hopping in the vertical and horizontal direction
by introducing the corresponding hopping intensities $t_v >0$ and $t_h >0$,
and the anisotropy parameter $\gamma$ such that
$t_v=\sqrt{\gamma}t_h$, with $0\leq \gamma \leq1$ \cite{DJ4}
(this somewhat unusual definition of the anisotropy parameter
is dictated by the simplicity of the expression for the effective interaction).
Then, in the case of the hole-particle invariant system,
the effective interaction (in the units of $U$) between the ions, up to order 4,  reads:
\begin{eqnarray}
\label{Expfmn}
& & \left[ \frac{t^{2}}{4}- \frac{3t^{4}}{16}-\frac{3}{8}\gamma
t^{4} + \frac{W}{8} \right] \sum\limits_{\langle x,y \rangle_{1,h}}
s_{x}s_{y} +\nonumber \\
& & \left[ \gamma \frac{t^{2}}{4} -\frac{3}{8}\gamma t^{4}
- \gamma^{2}\frac{3t^{4}}{16} + \frac{W}{8} \right]
\sum\limits_{\langle x,y \rangle_{1,v}} s_{x}s_{y} +
\nonumber \\
& & \left[ \gamma \frac{3t^{4}}{16}-\frac{\tilde{\varepsilon}}{16}
\right] \sum\limits_{\langle x,y \rangle_{2}} s_{x}s_{y} +
\frac{t^{4}}{8} \sum\limits_{\langle x,y \rangle_{3,h}}
s_{x}s_{y} +
\nonumber
\\
& & \gamma^{2} \frac{t^{4}}{8} \sum\limits_{\langle x,y
\rangle_{3,v}} s_{x}s_{y} + \gamma \frac{t^{4}}{16} \sum\limits_{P}
\left(1+5s_{P}\right),
\end{eqnarray}
where $t=t_{h}/U$, $\langle x,y \rangle_{n,h}$ ($\langle x,y
\rangle_{n,v}$) stands for pairs of $n$th-order nearest-neighbor
sites (\nn{}-1st order, etc) in horizontal (vertical) directions,
and $s_P$ denotes a product of four spins whose sites constitute an
elementary square, $P$, on the lattice.

Since we are working with a truncated effective interaction,
we have to assign an order to the deviation of the
anisotropy parameter $\gamma$ from the value $1$ (the isotropic case).
For this purpose we define the anisotropy order, $a$,
and the new anisotropy parameter, $\beta_{a}$: $\gamma=1-\beta_{a}t^{a}$,
$\beta_{a} > 0$.
In order to analyse anisotropy effects with the effective interaction
truncated at certain order,
the anisotropy order has to be suitably adjusted.
In our work, we analyze the effective interaction truncated at 4th order,
hence the weakest admissible deviation from the isotropic case corresponds
to $a=2$.

\section{Ground-state phase diagrams}

\begin{figure}[p]
\begin{center}
\includegraphics[height=0.25\textheight]{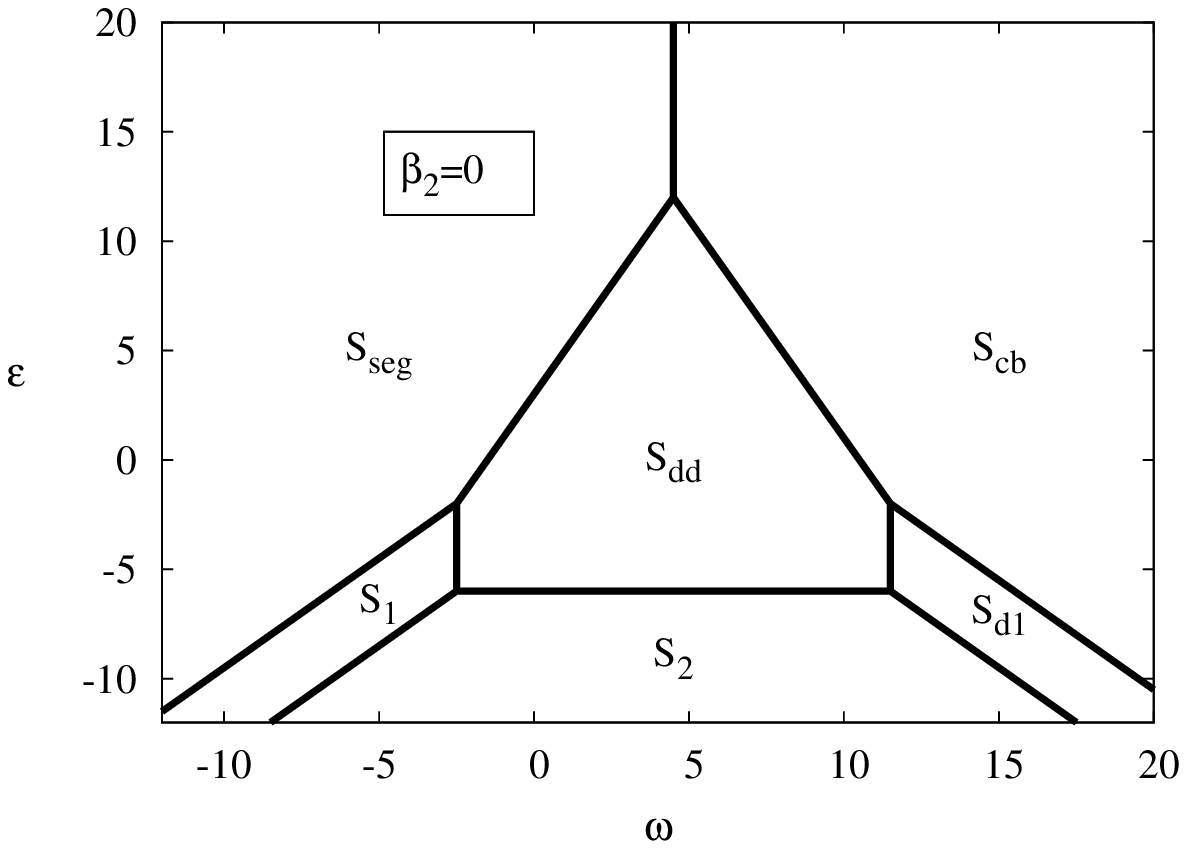}\\
\includegraphics[height=0.25\textheight]{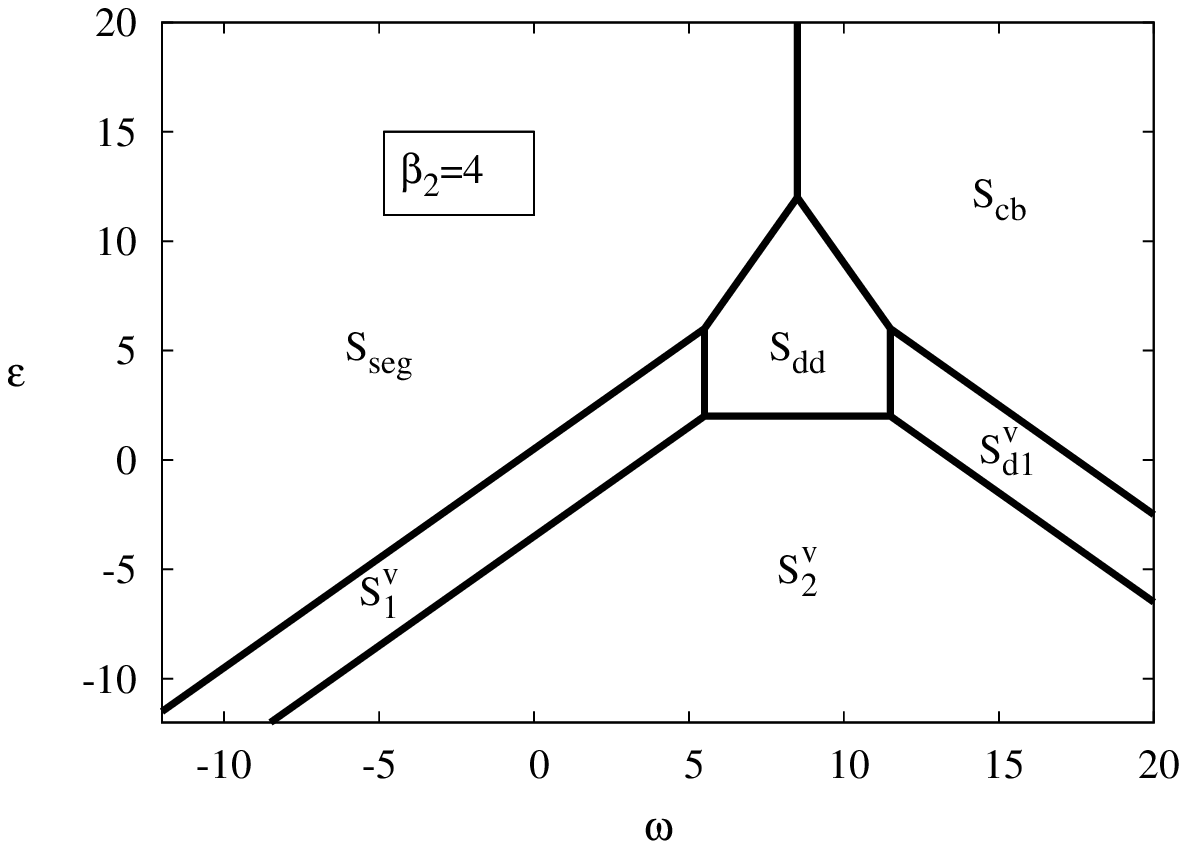}\\
\includegraphics[height=0.25\textheight]{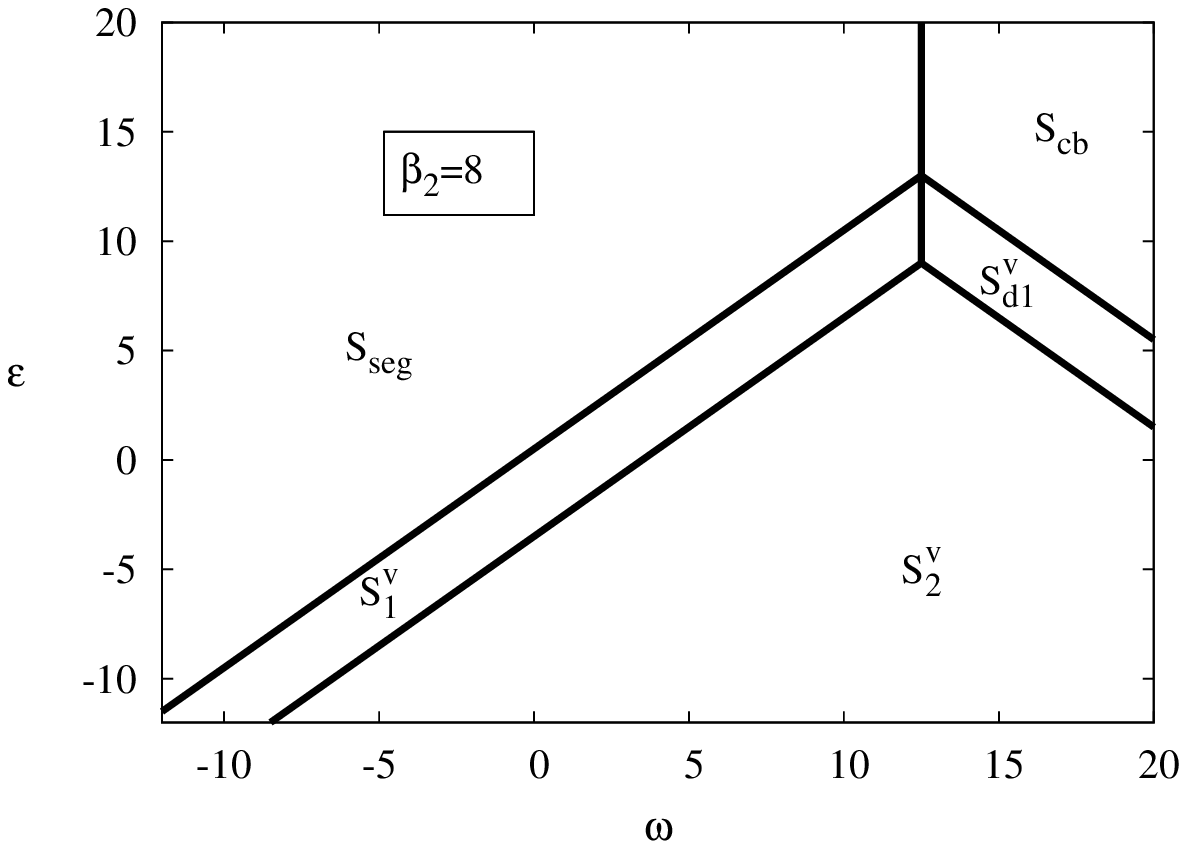}
\end{center}
\vspace{-3mm} \caption{Ground-state phase diagrams in the case of
the hole-particle invariant system, according to the effective interaction
(\ref{Expfmn}). The representative ion configurations of
the displayed phases are shown in Fig.~\ref{config}. The superscript $v$ by the
symbols of phases means that only phases with vertically oriented configurations
of ions are stable. ${\mathcal{S}}_{seg}$ stands for the segregated phase,
which is a mixture of two uniform phases: ${\mathcal{S}}_{+}$ (where $s_x=1$ at every site $x$) and
${\mathcal{S}}_{-}$ (where $s_x=-1$ at every site $x$).} \label{phd}
\end{figure}

The ground-state phase diagrams constructed with effective
interaction  (\ref{Expfmn}) are shown in Fig.~\ref{phd}. The
essential point is that they can also be thought of as ground-state
phase diagrams of the full quantum system, provided that the
coexistence lines of the phases are interpreted as strips whose
width is of the order $O(|t/U|^2)$. Inside these strips, we cannot
claim the stability of any phase, but we can exclude the stability
of some phases. Studies carried out by approximate methods  suggest
that these strips may accommodate infinitely many phases
\cite{GUJ,GJL,WL}. To unveil these phases rigorously, it is
necessary to construct phase diagrams according to the effective
interactions truncated at higher orders, but this task becomes
quickly hardly feasible.

Because of the hole-particle invariance, in all the phases of the phase diagrams
shown in Fig.~\ref{phd} the densities of the ions and the electrons are $1/2$;
off the hole-particle symmetry case, many other phases of different than $1/2$
densities can be proven to be stable \cite{DJ4}.
Suppose that initially our system is in a crystalline checkerboard phase,
$\mathcal{S}_{cb}$, Fig.~\ref{phd}.
Then, by decreasing the ``van der Waals interaction'' coupling $\omega$,
the system is always driven out of the crystalline state. However,
the terminal state and the way it is attained depend strongly on
variation of the tiny coupling controlled by $\varepsilon$.
It is worth to note here that the complete \nnn{}, 4th order,
effective interaction consists of a contribution from $H_{FK}$ and from $H_{vdW}$;
it is attractive if $\varepsilon > 3$, irrespectively of the value of the electron
hopping anisotropy controlled by $\beta_{2}$, and then it stabilizes the checkerboard
phase.
\begin{figure}[h]
\begin{center}
\includegraphics[width=0.6\textwidth]{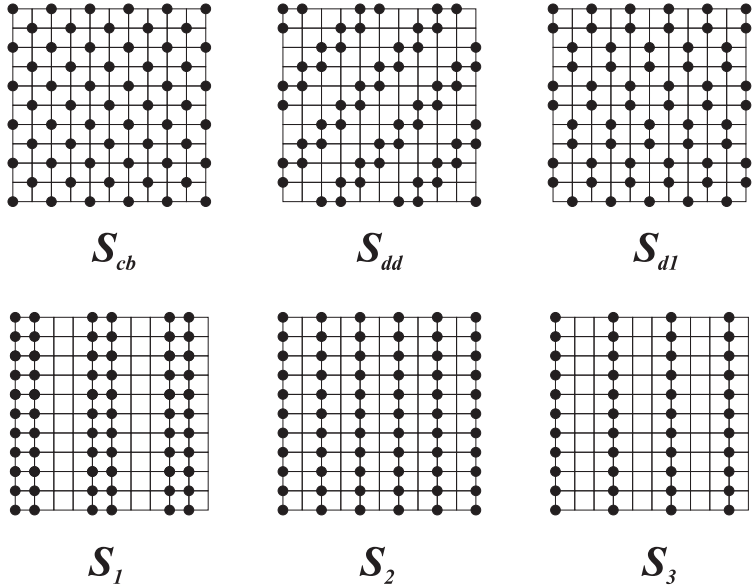}
\end{center}
\caption{Representative ion configurations of the phases considered in the paper.
${\mathcal{S}}_{cb}$  and ${\mathcal{S}}_{d1}$ are examples of checkerboard-like crystals,
${\mathcal{S}}_{dd}$  -- of diagonal-stripe phases, ${\mathcal{S}}_{1}$, ${\mathcal{S}}_{2}$,
${\mathcal{S}}_{3}$ -- of axial-stripe phases.
${\mathcal{S}}_{3}$ is absent in the phase diagrams of Fig.~\ref{phd}.}
\label{config}
\end{figure}
Concerning a terminal state, one can observe that if $\varepsilon$
does not decrease (while $\omega$ decreases), then the terminal
state is always the segregated phase, which is, in this case, a fifty-fifty
mixture of the uniform phases $\mathcal{S}_{+}$ and
$\mathcal{S}_{-}$. If, on the other hand, $\varepsilon$ does not
increase, then whether the segregated phase is attained or not
depends on the rate of its variation as compared with that of
$\omega$. In particular, if $\varepsilon$ decreases sufficiently
fast, then the process of deterioration of the crystal terminates in
one of the axial-stripe phases $\mathcal{S}_{1}$,
$\mathcal{S}_{1}^v$, $\mathcal{S}_{2}$, $\mathcal{S}_{2}^v$. The
state of segregation is not attained.
This is reminiscent of the
results of qualitative analyses of classical lattice gases with
competing interactions, repulsive long-range dipole-dipole or Coulomb interactions
and attractive short-range interactions \cite{BMWB,JKS,SK}.

Now, concerning the sequence of intermediate phases visited by the system before
the terminal state is attained,
we can see in Fig.~\ref{phd} only a few elements of this sequence
(see the remark above concerning the coexistence lines). Apparently, two cases can be realized.
In the first one, the only intermediate phase is a mixture of
$\mathcal{S}_{cb}$, $\mathcal{S}_{+}$, and $\mathcal{S}_{-}$.
This occurs, for instance, if the value of $\varepsilon$ in the initial state is large enough
and is kept constant. In this case, the system being initially in a crystalline state
attains the terminal segregated phase directly via a 1st order phase transition
\cite{DJ2,DJ3}.
In the second case, before a terminal state (the segregated phase or an axial-stripe phase)
is attained, the system visits numerous intermediate phases; at the end of this sequence
one always finds axial-stripe phases. This case is realized, for instance,
if $\varepsilon$ is sufficiently small and kept constant.
Then, the \nnn{}, 4th order, effective interaction
becomes repulsive and frustrates the interactions stabilizing the checkerboard phase.
As a result, as $\omega$ decreases, first the ions form dimers which are distributed
in a checkerboard-like manner in the phase $S_{d1}$, then the the particles form extended
strings, i.e. completely filled lattice lines, like in the axial-stripe phases,
$S_{1}$ and $S_{3}$.
The latter succession of phases is reminiscent of those observed in many other systems
with competing interactions (see for instance \cite{BMWB,Stroud} for results
of computer simulations and \cite{giuliani123} for rigorous results) and conforms to
the following observation:
purely repulsive interactions tend to proliferate as small as possible
aggregates of particles and arrange them in checkerboard-like crystalline patterns,
while attractive interactions tend to make the aggregates of
particles as large as possible, which results eventually in segregated-like phases
with mesoscopic regions completely filled with particles or completely empty.
Let us emphasize again that, when a transition from the checkerboard crystal to a
segregated phase or a striped phase is accomplished, the sequence of visited phases is,
most probably, much longer. The phases visible in the diagrams of Fig.~\ref{phd},
being a part of this sequence,
point out only to some tendencies. Apparently, in a process of deterioration of a crystal,
striped phase constitute a stage just preceding a transition to a segregated
phase.

Finally, let us consider an impact of electron hopping anisotropy on the phase diagrams.
For definiteness,  we set the vertical hopping weaker than the horizontal one
(i.e. $\beta_2 >0$). Any hopping anisotropy breaks the rotational symmetry of the
system and out of the phases that are not invariant with respect to all rotations,
select some with a specified orientation.
For instance, among the axial-stripe phases it stabilizes those stripes that are
oriented in the direction of a weaker hopping (vertical in our case).
This effect is clearly visible in the phase diagrams of Fig.~\ref{phd},
where for a nonzero anisotropy of \nn{} hopping only vertically oriented striped
phases remain stable.
Let us emphasize that this property holds
not only for a truncated effective interaction for which the phase diagrams are
actually constructed but for the full quantum system as well.
In \cite{Derzhko}, it was demonstrated that an analogous effect occurs also
for diagonal-stripe phases in the presence of an anisotropy of \nnn{} electron hoppings.

The effect of orienting axial-stripe phases in the direction of a weaker hopping
has been found also in studies of a Hubbard model in the framework of real-space
Hartree-Fock approximation \cite{oles}.

\section{Specific properties of axial-stripe phases}

All the conclusions of the previous section are based on a rigorous
analysis of ground-state phase diagrams of our system, for
sufficiently large electron-ion coupling and for a half-filled
system, with the main result referring to the stability of striped
phases. In this section, we look into properties of the axial-stripe
phases to see how important are the above conditions for the drawn
conclusions, whether the effect of orienting axial stripes in the
direction of weaker hopping, seen in the diagrams of previous
section, might persist beyond the regimes of strong coupling and
half-filling. Moreover, it should be interesting to unveil the
electronic properties of axial-stripe phases, which have been left
untouched by the analysis of previous section.

Numerical calculations of the thermodynamic functions considered below are based on
exact diagonalization of the electron subsystem in axial striped phases
$\mathcal{S}_1$, $\mathcal{S}_2$, and $\mathcal{S}_3$ (the corresponding spectra and some of their
properties are reproduced in Appendix). In this section we present only some properties, verified
for the phases $\mathcal{S}_1$, $\mathcal{S}_2$, and $\mathcal{S}_3$ (see Fig.~\ref{config};
the lack of subindex ``v'' indicates that we consider both the horizontally and vertically
oriented stripes). We think that the presented properties are characteristic for the whole family of
the axial-stripe phases. To support our conclusions, we display only plots of discussed quantities for
the phase $\mathcal{S}_3$ (see Fig.~\ref{config}), which is not present in the displayed here hole-particle
symmetric phase diagrams (it is visible in diagrams off the hole-particle symmetry \cite{DJ3,DJ4}),
but appears to be a good representative of the family of axial-stripe phases.

\begin{figure}[h]
\begin{center}
\includegraphics[width=0.49\textwidth]{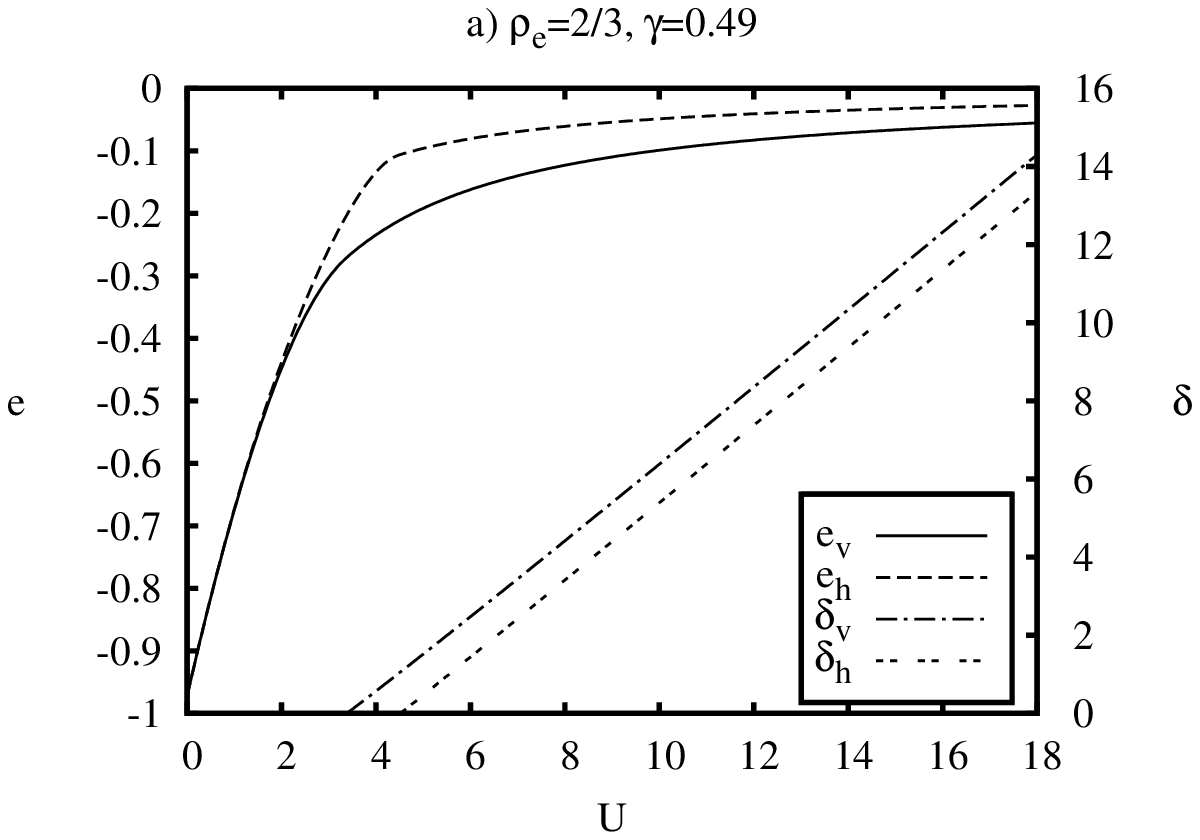}
\includegraphics[width=0.49\textwidth]{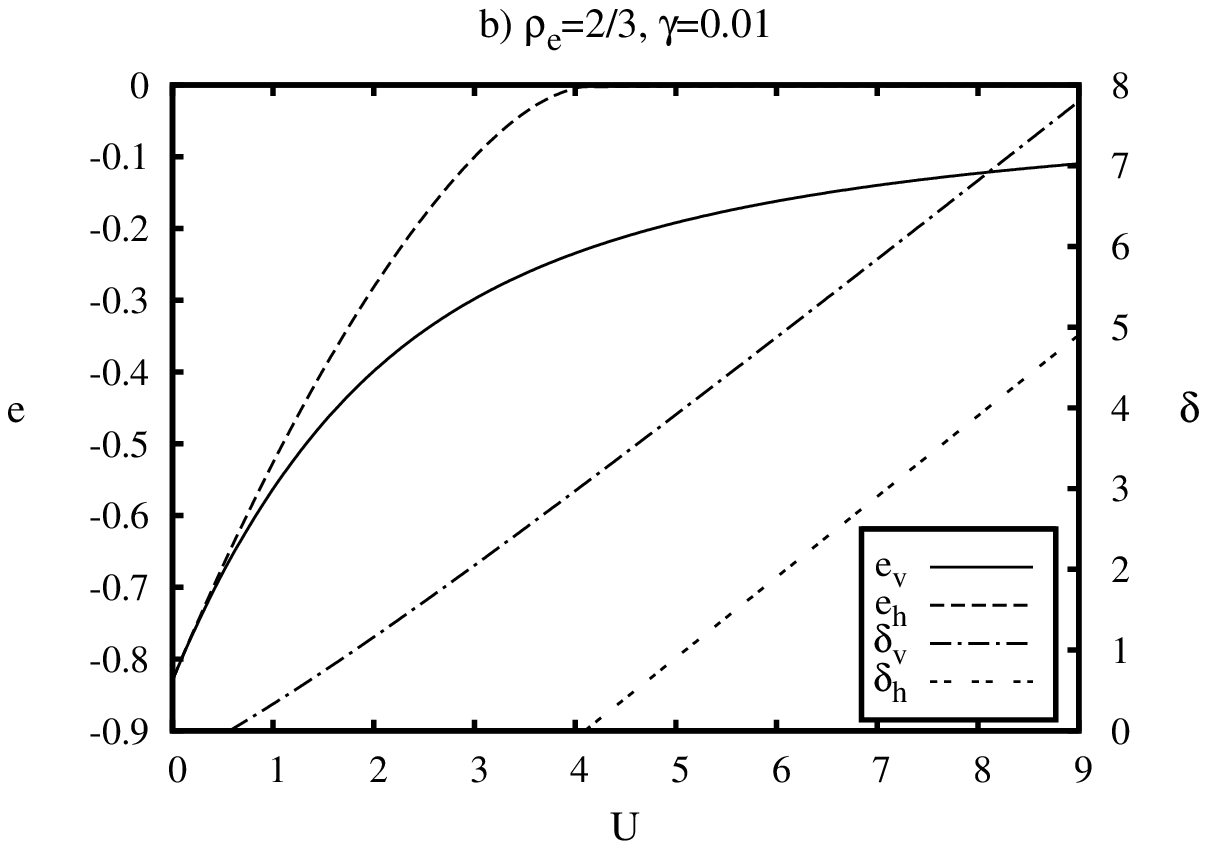}
\end{center}
\caption{Energy density (per particle),  $e$, and the width,
$\delta$, of the gap at the Fermi level in the half-filled phases
$\mathcal{S}_{3}^{v}$ and $\mathcal{S}_{3}^{h}$, as functions of
$U$, for two values of the anisotropy parameter $\gamma$.}
\label{hfgse}
\end{figure}

We start with comparing the ground-state internal-energy densities of vertically and horizontally
oriented axial-stripe phases of a half-filled system, for a small and a large electron-hopping
anisotropy, see Fig.~\ref{hfgse}. Let us recall that according to the phase diagrams displayed
in previous section, it is the phase whose stripes  are oriented in the direction of weaker hopping
(vertical in our case) that has a lower energy. This effect occurs in a situation, where
all the phases in the phase diagrams constructed with the method used in our work
(strong-coupling expansion of half-filled system) are insulating.
The reason is that for $U$ so large that the expansion is convergent,
a gap opens at the Fermi level of any stable periodic phase \cite{GM}.

Now, according to our calculations carried out for axial stripes,
the vertically oriented stripes seem to have lower energy than those
perpendicular to this direction, for all values of $U > 0$.
Of course, we cannot exclude numerically that
this holds only above some positive value of $U$, but in any case
this value is much smaller than the values of $U$ admissible in
the expansions referred to in Section 2.
It follows from dispersion relations (reproduced in Appendix) that for sufficiently large $U$,
in both systems, of vertical and horizontal stripes, there is a gap at the Fermi level.
This gap opens for smaller values of $U$ and is larger for vertical
stripes (those along the direction of weaker hopping), than the horizontal ones.
But the relative stability of vertical stripes persists even for the values
of $U$, for which there is no gap at the Fermi level.
The above observations shine more light on the stability of stripes oriented
in the direction of weaker hopping, relative to perpendicular stripes:
the effect found in the strong-coupling phase diagrams, when the discussed phases
are insulating, may persist also when they are metallic.

However, if we abandon the condition of half-filling the situation is different.
\begin{figure}[h]
\begin{center}
\includegraphics[width=0.49\textwidth]{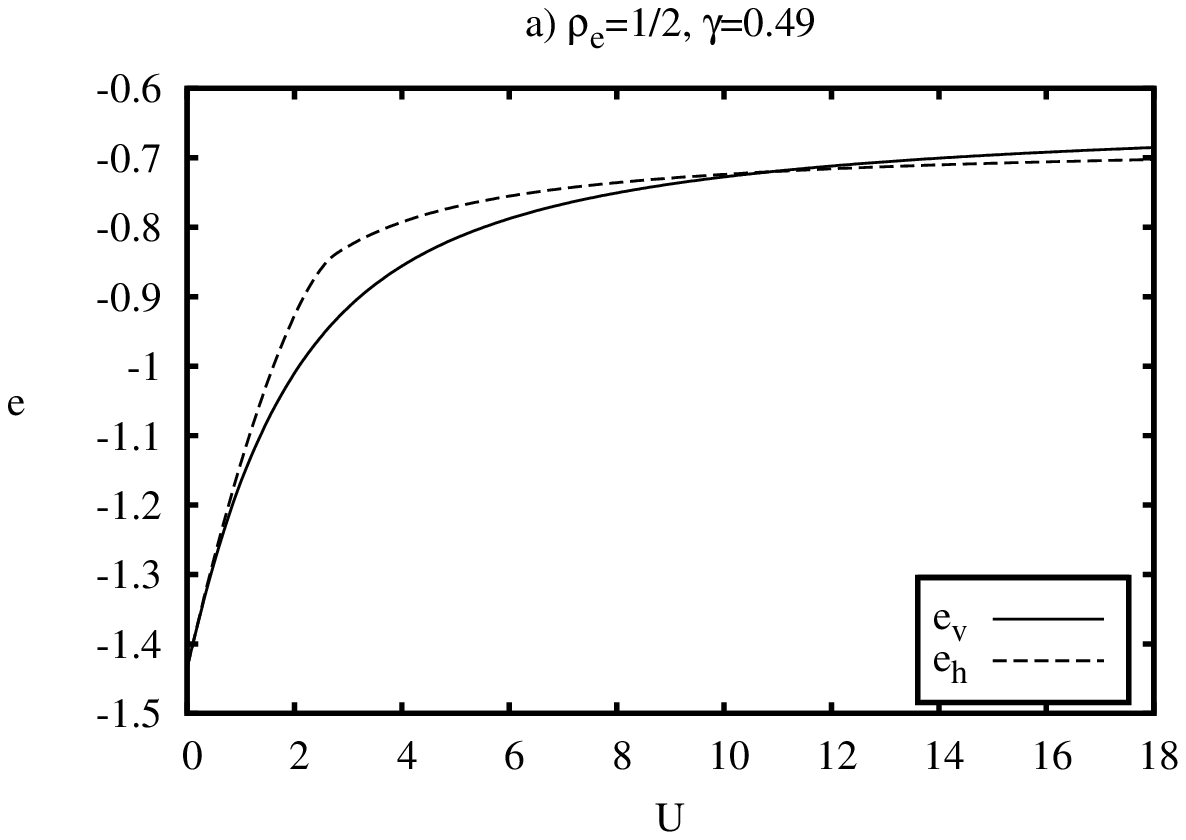}
\includegraphics[width=0.49\textwidth]{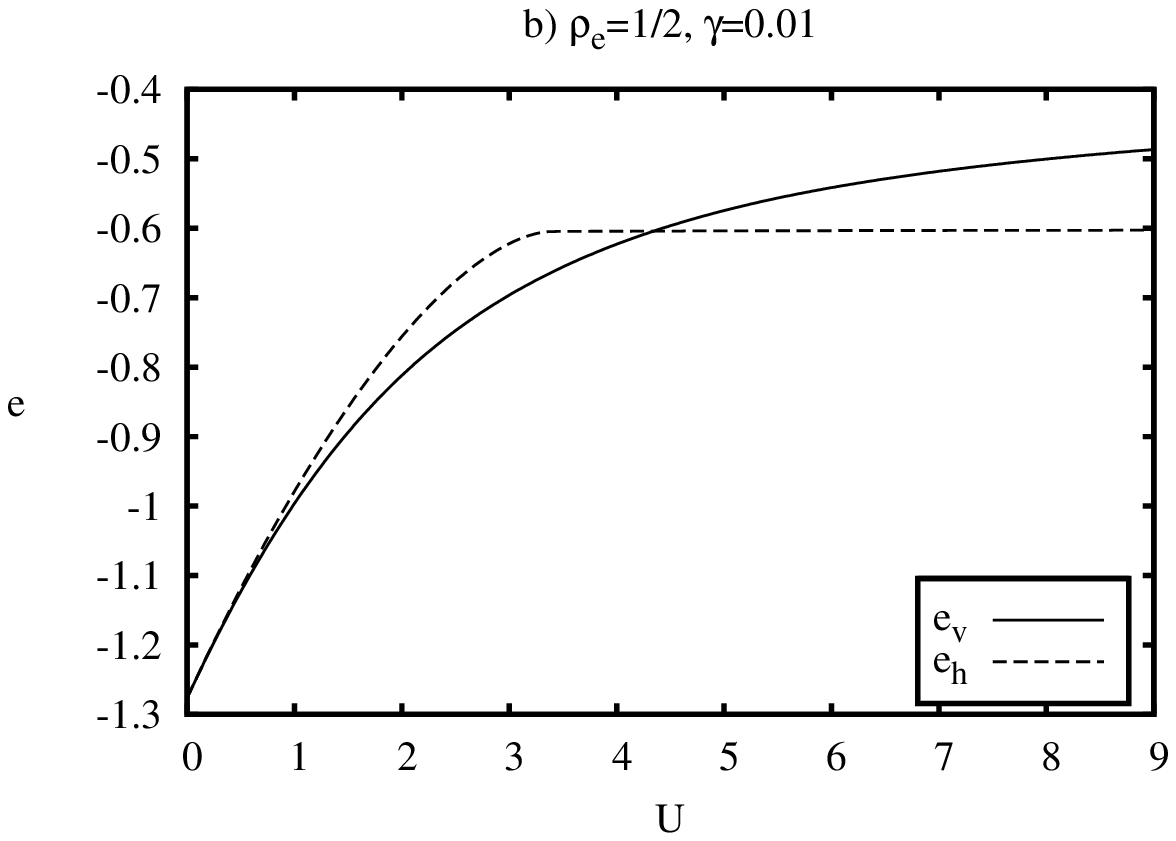}\\
\includegraphics[width=0.49\textwidth]{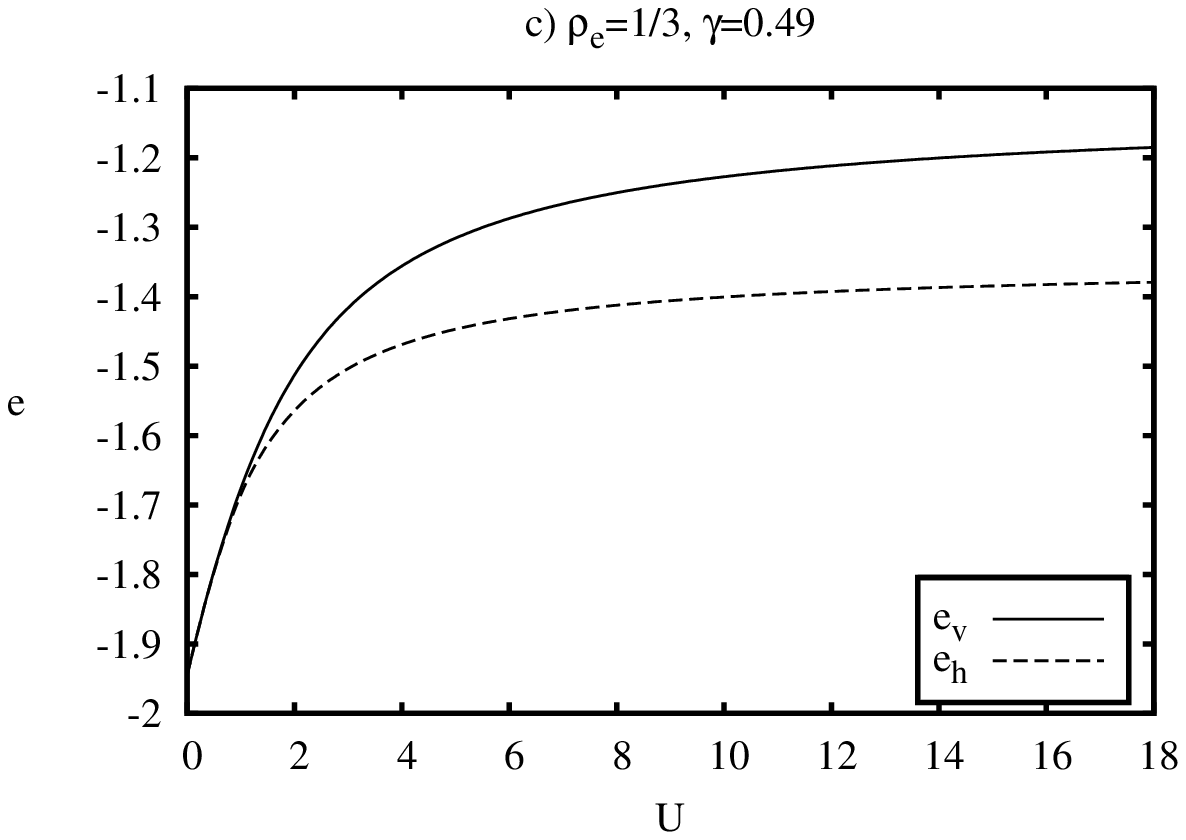}
\includegraphics[width=0.49\textwidth]{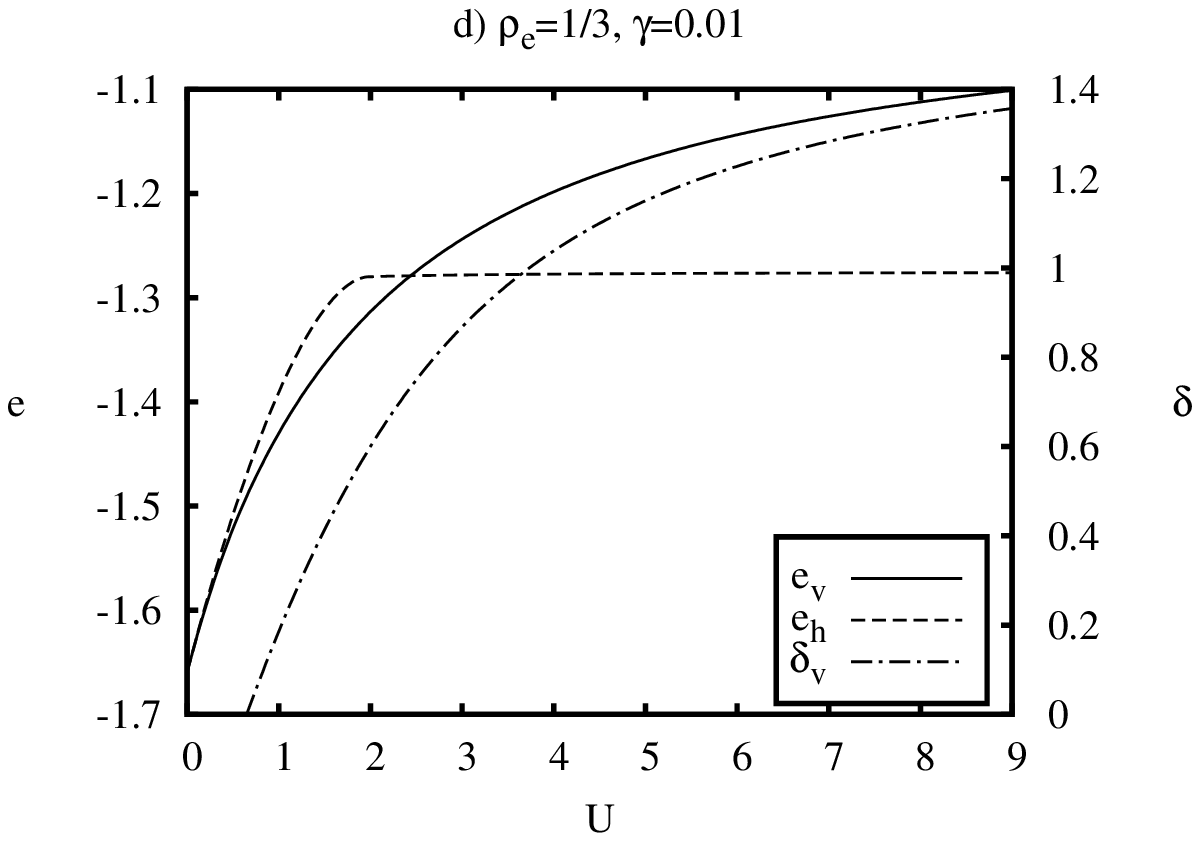}
\end{center}
\caption{Energy density (per particle), $e$, in the phases
$\mathcal{S}_{3}^{v}$ and $\mathcal{S}_{3}^{h}$ for two values of
electron density off the half-filling, and for two values of the
anisotropy parameter $\gamma$. Only in case d) a gap of width
$\delta$ opens at the Fermi level of $\mathcal{S}_{3}^{v}$.}
\label{ohfgse}
\end{figure}
In Fig.~\ref{ohfgse} we display the plots of the ground-state internal-energy densities of vertical
and horizontal stripes, for a small and a large electron hopping anisotropy, and for electron
densities $\rho_e=1/2$ and $\rho_e=1/3$.
In the cases displayed in Fig.~\ref{ohfgse} a,b,c there is no gap at the Fermi level, whether the stripes are
vertical or horizontal, while in the case of Fig.~\ref{ohfgse}d the gap at the Fermi level opens for vertical
stripes (because of sufficiently strong hopping anisotropy).
Nevertheless, in all the cases there is a critical value of $U$ above which it is the horizontally
oriented phase that has lower energy. That is, for large $U$ but away of half-filling,
the vertical stripes exchange their stability with horizontal stripes.

Some of the spatial correlations properties of the electron subsystem  are determined by off-diagonal
matrix elements (in position basis $\{|{\bf r} \rangle \}$) of the one-body reduced density
operator $\hat{\rho}^{(1)}$ (hereafter called the correlation functions), which can be
expressed by the  eigenfunctions in position representation.
If the Fermi level is located at an upper edge of a band,
which does not overlap with the bands of higher energy, then in the limit of an infinite system,
the correlation function depending on two lattice positions ${\bf{r}}$ and ${\bf{r^{\prime}}}$
assumes the form
\begin{eqnarray*}
\langle {{\bf{r}}} | \hat{\rho}^{(1)} | {{\bf{r^{\prime}}}} \rangle
=\frac{1}{(2\pi)^{2}} \int\limits_{-\pi}^{\pi} dk_{h} \int\limits_{-\pi}^{\pi} dk_{v} K({\bf{r}},{\bf{k}}) K^{*}({\bf{r^{\prime}}},{\bf{k}})
\exp{\left( i{\bf{k}}({\bf{r}}-{\bf{r^{\prime}}}) \right)},\\
\end{eqnarray*}
where the function $K({\bf{r}},{\bf{k}})$ is given in terms of the components (in the plane-wave
basis) of eigenvectors corresponding to completely filled bands.
As a matter of fact, in the case of axial-stripe phases these components,
hence the function  $K({\bf{r}},{\bf{k}})$, do not depend on the wave
vector component in the direction of stripes (see Appendix).
Let us consider for definiteness the case of vertically
oriented stripes. Then, in the vertical direction, that is for ${\bf{r^{\prime}}}={\bf{r}}+m{\bf{j}}$,
with $m$ being an integer and ${\bf{j}}$  standing for a lattice translation vector in the vertical
direction, the correlation function can be written as
\begin{eqnarray*}
\frac{1}{(2\pi)^{2}} \int\limits_{-\pi}^{\pi}dk_{h} K({\bf{r}},k_{h}) K^{*}({\bf{r^{\prime}}},k_{h})
\int\limits_{-\pi}^{\pi}dk_{v} \exp{\left( imk_{v} \right)},
\end{eqnarray*}
where the  last integral vanishes identically.

\begin{figure}[h]
\begin{center}
\includegraphics[width=0.49\textwidth]{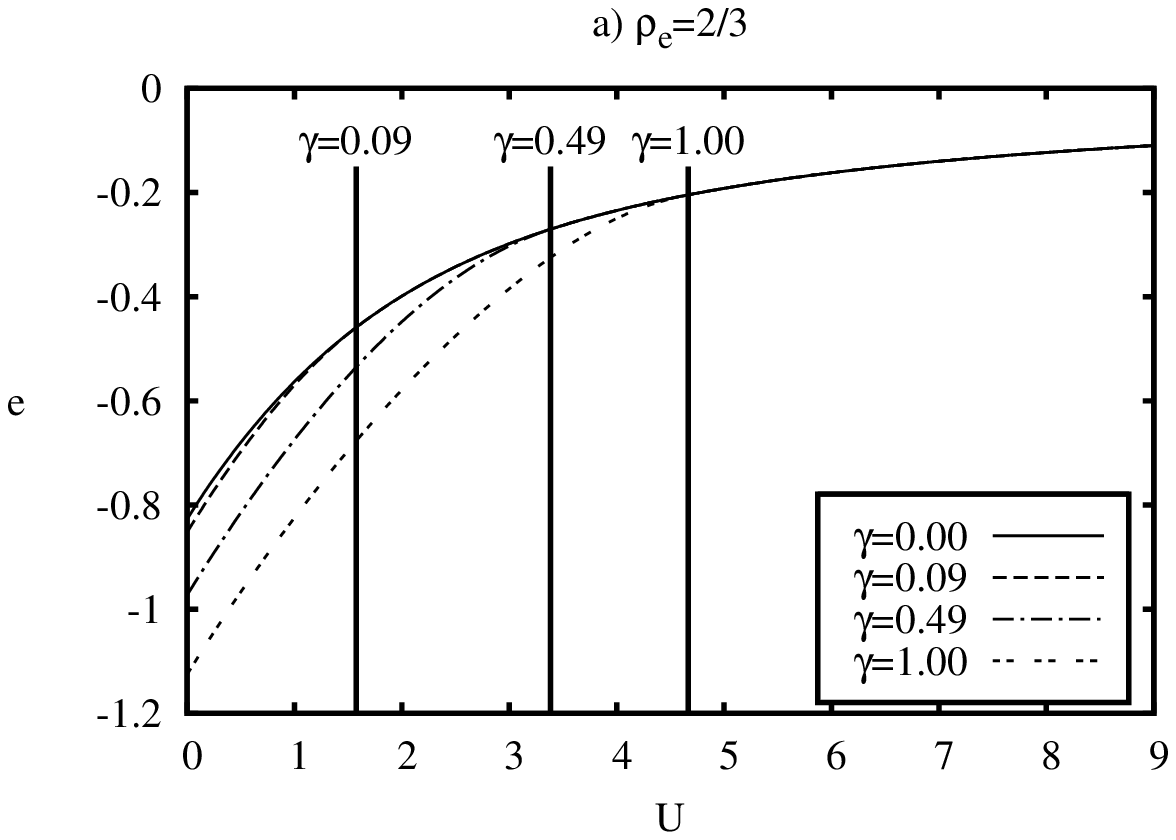}
\includegraphics[width=0.49\textwidth]{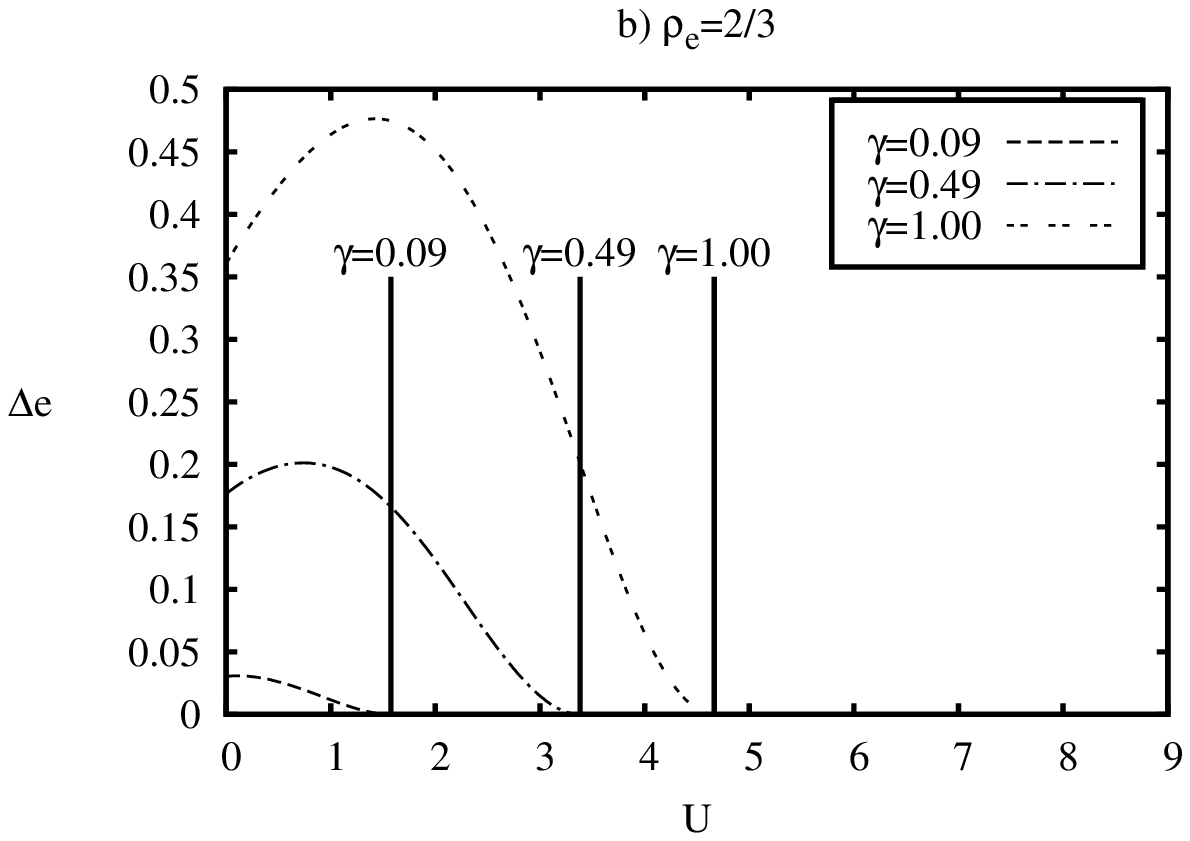}
\end{center}
\caption{Energy density (per particle), $e$, and relative energy
density, $\Delta e=|(e(\gamma)-e(0))/e(0)|$, for three nonzero
values of the anisotropy parameter $\gamma$, in the phase
$\mathcal{S}_{3}^{v}$. The vertical lines are located at the values
of $U$, where the gap opens. The highest (continuous) curve in part
a) corresponds to the 1D case.} \label{rele}
\end{figure}

Let us consider now half-filled axial-stripe phases for fixed $U$.
A sufficiently strong anisotropy (a sufficiently weak hopping along the stripes) opens a gap
at the Fermi level,
and then the chains perpendicular to the stripes become independent.
Therefore, for sufficiently  large anisotropy the ground-state extensive thermodynamic quantities
of the whole 2D system become equal to those of 1D chains perpendicular to the stripes
({\ie} with zero hopping in the direction of stripes).
This effect is clearly visible in Fig.~\ref{rele}a, where for the half-filled, vertically oriented,
axial-stripe phase $\mathcal{S}^{v}_3$
we show the plots of ground-state internal energy densities versus $U$, for different anisotropies.
Starting from sufficiently large $U$, each plot of a 2D system
coincides with the plot for the corresponding 1D system.
To see more clearly that the
plots merge exactly at the value of $U$ for which the gap at the
Fermi level opens, we have plotted also the relative differences of
internal-energy densities $\Delta e$ (see Fig.~\ref{rele}b and its
caption).

\begin{figure}[h]
\begin{center}
\includegraphics[width=0.49\textwidth]{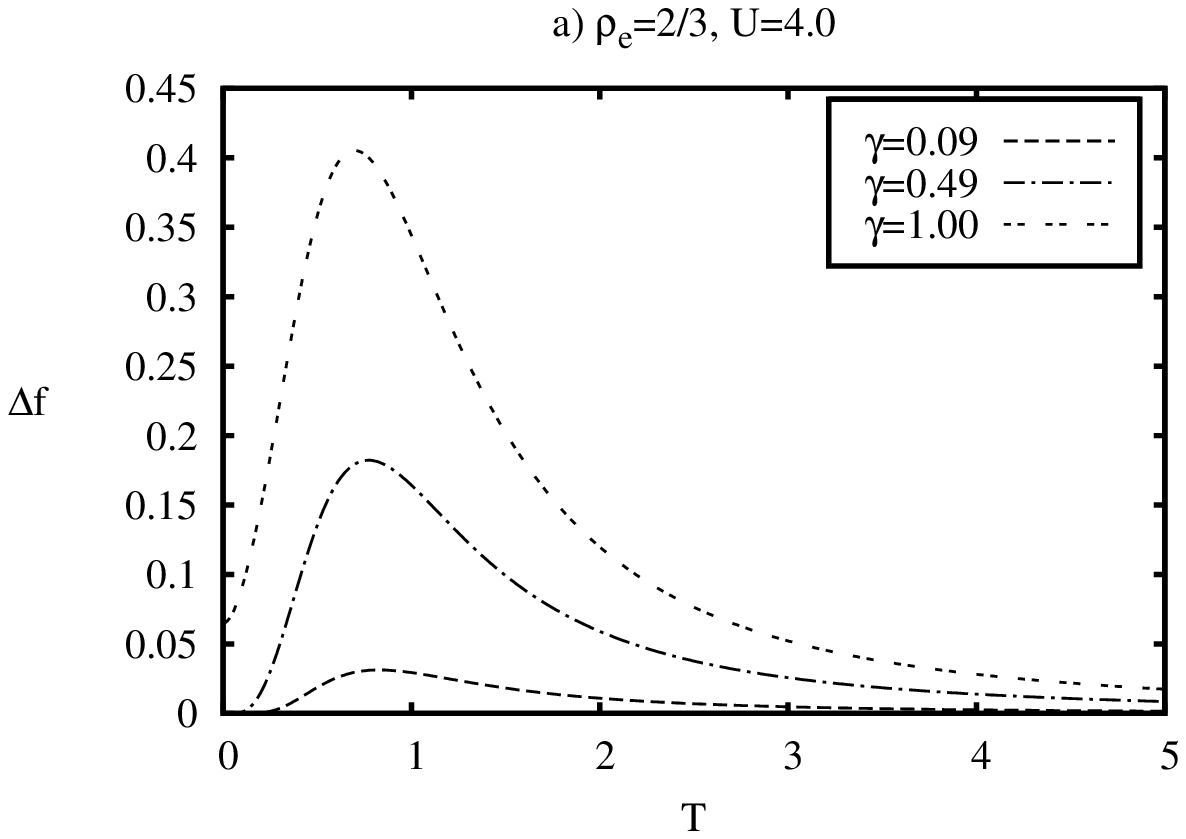}
\includegraphics[width=0.49\textwidth]{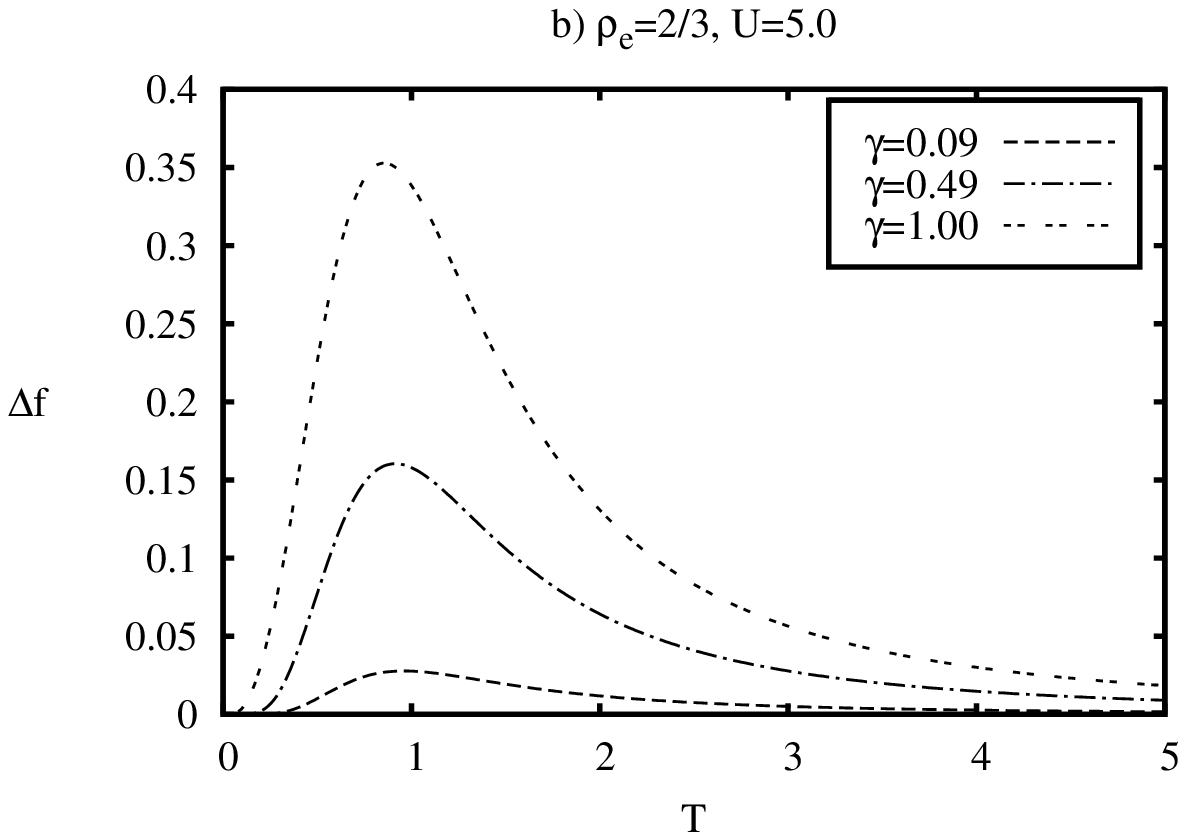}
\end{center}
\caption{Relative difference of free energy densities, $\Delta
f=|(f(\gamma)-f(0))/f(0)|$, for three values of the anisotropy
parameter $\gamma$, in the phase $\mathcal{S}_{3}^{v}$ with two
values of $U$.} \label{relf}
\end{figure}

It is quite clear that thermal fluctuations will destroy the ideal
independence of chains at zero temperature. However, it is
interesting to learn how effective they are, by calculating,
for instance, the Helmholtz free-energy-density relative differences,
$\Delta f$, versus temperature $T$ (see Fig.~\ref{relf}a,b and its
caption). We display the corresponding plots for three different
anisotropies and two values of $U$.
Let us note that the larger is $U$ and the stronger is the
anisotropy the larger is the gap at the Fermi level in stripes
oriented in the direction of the weaker hopping. For the smaller of
the $U$ values (Fig.~\ref{relf}a), the gap opens only for
sufficiently strong anisotropy. In the isotropic case (the highest
curve) there is no gap,  the chains perpendicular to the stripes are
dependent even at $T=0$, so the plot of the corresponding  $\Delta
f$ starts above zero. For the larger of the $U$ values
(Fig.~\ref{relf}b) there is a gap for any anisotropy and all the
curves start at zero. It is apparent that for a specified tolerance
for the deviation $\Delta f$ (of a few per cent of the maximal
value), the larger is the gap the higher is the temperature above
which this tolerance is exceeded. Below this temperature the chains
can be considered as approximately independent.

\section{Summary}
We have studied an extension of a microscopic quantum model of
crystallization, proposed by Kennedy and Lieb, in which light hopping
electrons interact on-site only with heavy immobile ions. The original
model has been extended by a weak short-range attractive interaction between
ions, whose purpose is to mimic the effect of van der Waals forces.
In the framework of the extended model,
the stability of striped phases has been analyzed on a
rigorous basis. Moreover, we have looked closely into properties of
a special class of striped phases, the axial-stripe phases, particularly
into properties of their electron subsystem under presence of a hopping
anisotropy. It is known already that an anisotropy of the electron
hopping has a strong impact on the properties of electron systems
\cite{JK3}; even a half-filled system of free electrons in a
constant potential (no gap at the Fermi level) turns into insulator,
in the direction of a weaker hopping, as soon as a hopping anisotropy
is ``switched on''. This is a consequence of an exponential decay of
correlations in this direction, for any nonzero anisotropy. In the
model considered here, the effects of hopping anisotropy are even
more striking. We have demonstrated that for sufficiently large
anisotropy (depending on the value of $U$), the stable axial stripes
(i.e. oriented in the direction of a weaker hopping) decouple into 1D
chains, which are perpendicular to the stripes. The reason is that
(for a given $U$) a sufficiently strong anisotropy opens a gap at
the Fermi level, and this in turn leads to vanishing of electron
correlations along the stripes. Consequently, while the ion
subsystem develops a 2D long-range order, the electron subsystem may
develop only a 1D long-range order, and the compound 2D system
behaves like a collection of chains. It is tempting to suggest that
an analogous effect may occur in many other systems developing
stripes, observed experimentally or studied theoretically, some of
which were mentioned in the Introduction. The appearance of stripes
in some degrees of freedom may signal a significant reduction of
correlations along the stripes among other degrees of freedom of a
compound system.

\section{Appendix}

Here we present some spectral quantities for electrons in periodic
potentials given by vertical ion configurations
${\mathcal{S}}^{v}_{2}$, ${\mathcal{S}}^{v}_{3}$ and
${\mathcal{S}}^{v}_{1}$, under periodic boundary conditions. We
denote the dispersion relations as $\lambda_{m}^{(l)}({\bf k})$ and
the gaps widths as $\delta_{m}^{(l)}$, where $m$ labels
configurations ($m=1,2,3$) and $l$ counts the bands and gaps from
bottom to top. The corresponding quantities for horizontal
configurations can be obtained by exchanging $t_{v}$ and $t_{h}$ in
the formulae below.

First, for all the considered cases we define $\varepsilon_{v}$ as
\begin{eqnarray*}
\varepsilon_{v}=2t_{v}\cos{\left( k_{v} \right)}.
\end{eqnarray*}

Then, for ${\mathcal{S}}^{v}_{2}$
\begin{eqnarray*}
\lambda^{(l)}_{2}({\bf{k}})&=&\varepsilon_{v}+u\pm \Delta_{h},\\
\delta_{2}&=&\left| U \right| -4\left| t_{v} \right|,
\end{eqnarray*}
where $\varepsilon_{h}=2t_{h}\cos{\left( k_{h}/2 \right)}$, $u=U/2$
and $\Delta_{h}=\sqrt{\varepsilon_{h}^{2}+u^{2}}$;

for ${\mathcal{S}}^{v}_{3}$
\begin{eqnarray*}
\lambda^{(l)}_{3}({\bf{k}})&=&\varepsilon_{v}+u
+2\sqrt{t_{h}^{2}+u^{2}}
\cos{\left( \frac{ \phi_{h}}{3}+\frac{2\pi l}{3} \right)}, \\
&& l=1,2,3;\\
\delta^{(l)}_{3}&=& -4\left| t_{v} \right|+ 2\sqrt{3\left(
t_{h}^{2}+u^{2} \right)} \sin{\left( \dfrac{\phi^{*}_{h}}{3} +
\dfrac{2\pi (l-1)}{3} \right)}, \\
&&l=1,2;
\end{eqnarray*}
where $u=U/3$, $\phi_{h}=\arccos{ \left(
\dfrac{u^3+t_{h}^{3}\cos{(k_{h})}}{ \left( t_{h}^{2} + u^{2}
\right)^{3/2}}\right)}$ and $\phi^{*}_{h}=\arccos{\left(
\dfrac{u^3+(-1)^{l-1}|t^{3}_{h}|}{\left[ u^2+t_{h}^{2}
\right]^{3/2}} \right)}$;

for ${\mathcal{S}}^{v}_{1}$
\begin{eqnarray*}
\lambda^{(l)}_{1}({\bf{k}})&=&\varepsilon_{v}+u \pm
\sqrt{2t_{h}^{2}+u^{2} \pm 2|t_{h}|\Delta_{h}},\\
\delta_{1}^{(1)}&=&-4|t_{v}|+\sqrt{t_{h}^{2}+\left(|t_{h}|+|u|\right)^{2}}
-\sqrt{t_{h}^{2}+\left(|t_{h}|-|u|\right)^{2}},\\
\delta_{1}^{(2)}&=&-4|t_{v}|+2\left(\sqrt{t_{h}^{2}+u^{2}}-|t_{h}|\right),\\
\delta_{1}^{(3)}&=&\delta_{1}^{(1)};\\
\end{eqnarray*}
where $u=U/2$, $\Delta_{h}=\sqrt{t_{h}^{2}\cos^{2}{(k_{h}/2)}+u^2}$.

\begin{figure}[p]
\includegraphics[width=0.41\textwidth]{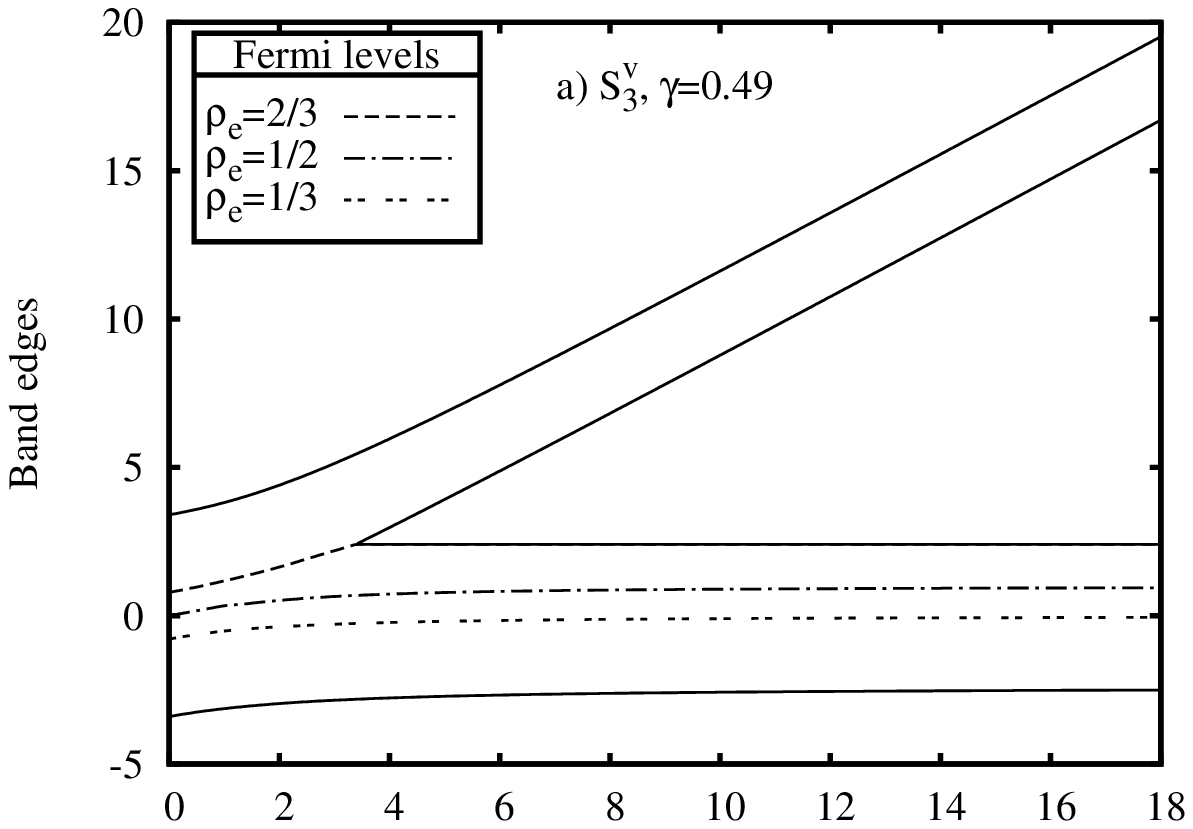}
\hfill
\includegraphics[width=0.41\textwidth]{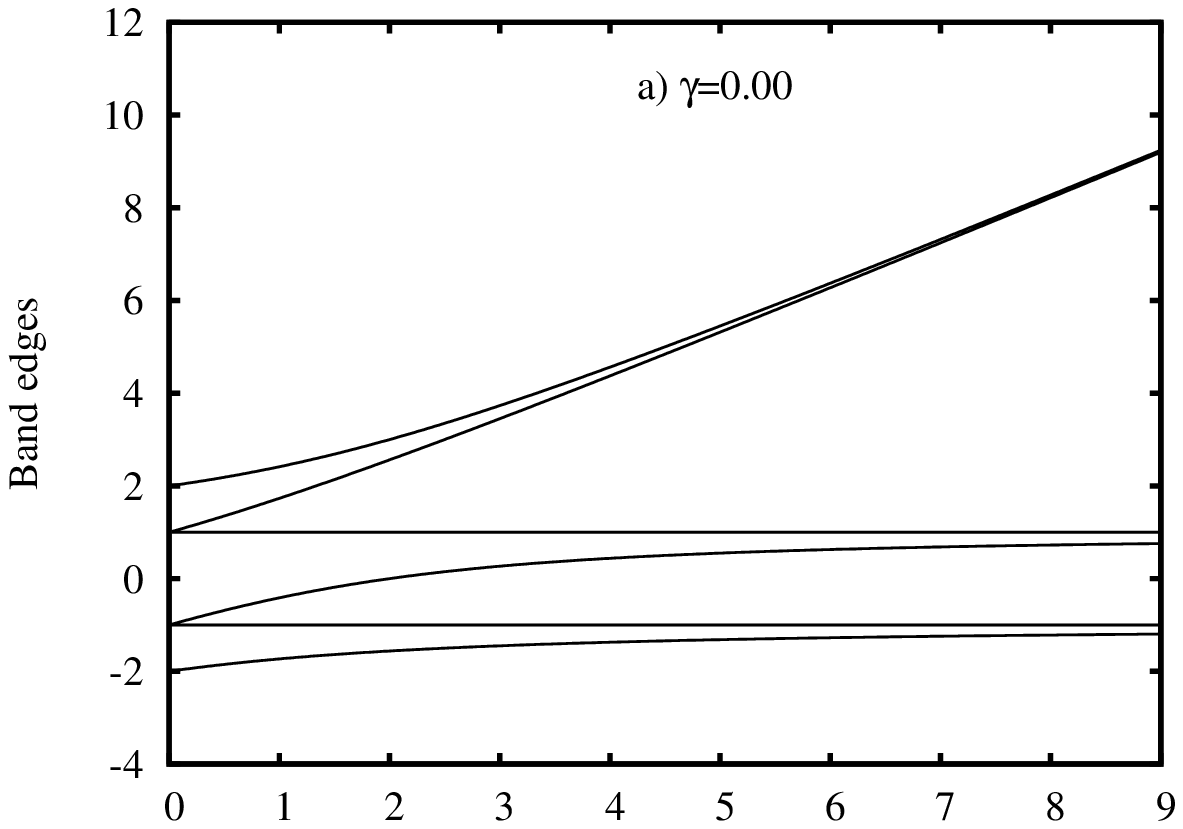}\\
\includegraphics[width=0.41\textwidth]{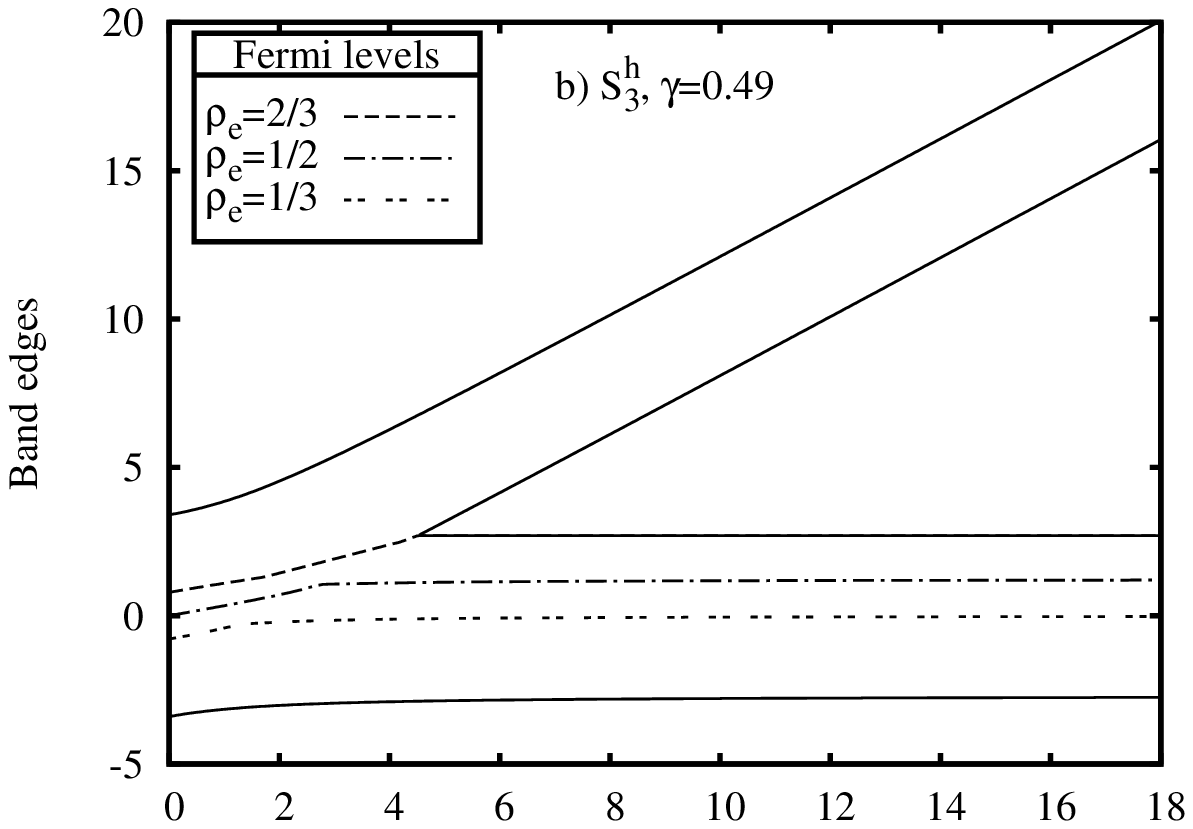}
\hfill
\includegraphics[width=0.41\textwidth]{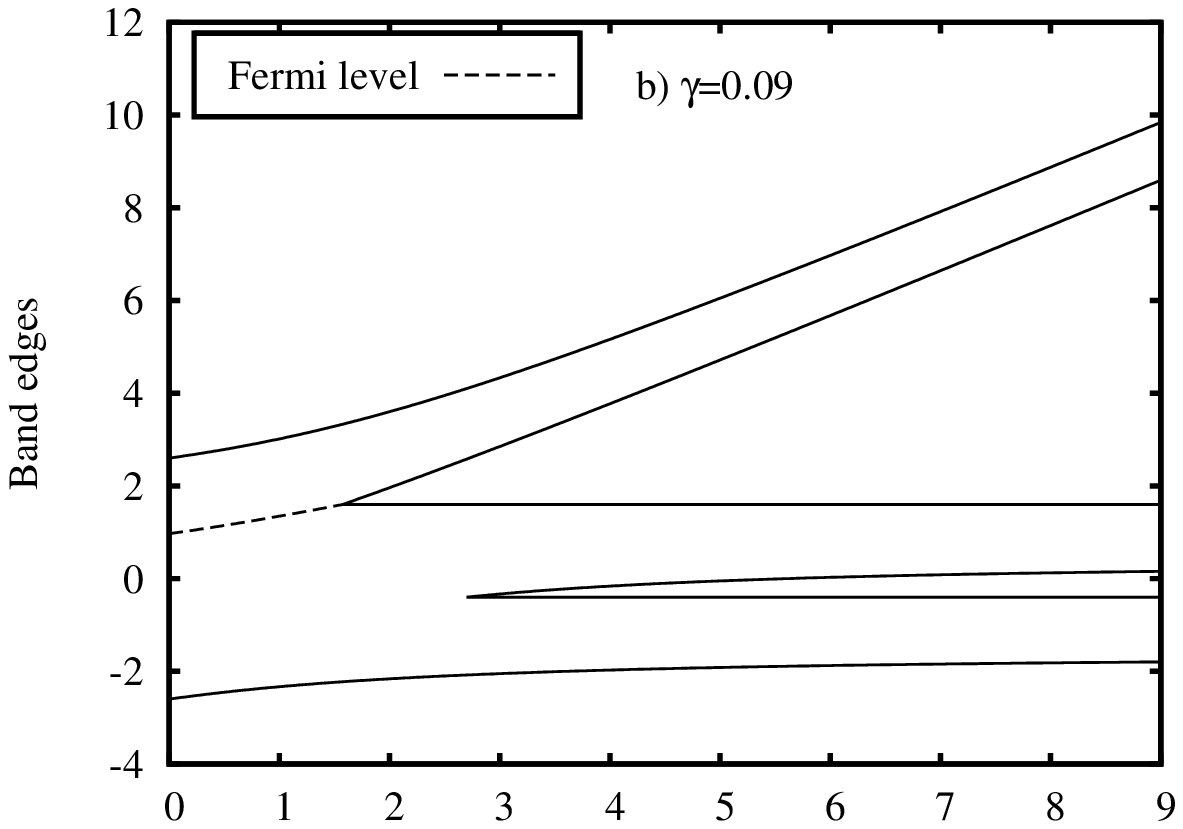}\\
\includegraphics[width=0.41\textwidth]{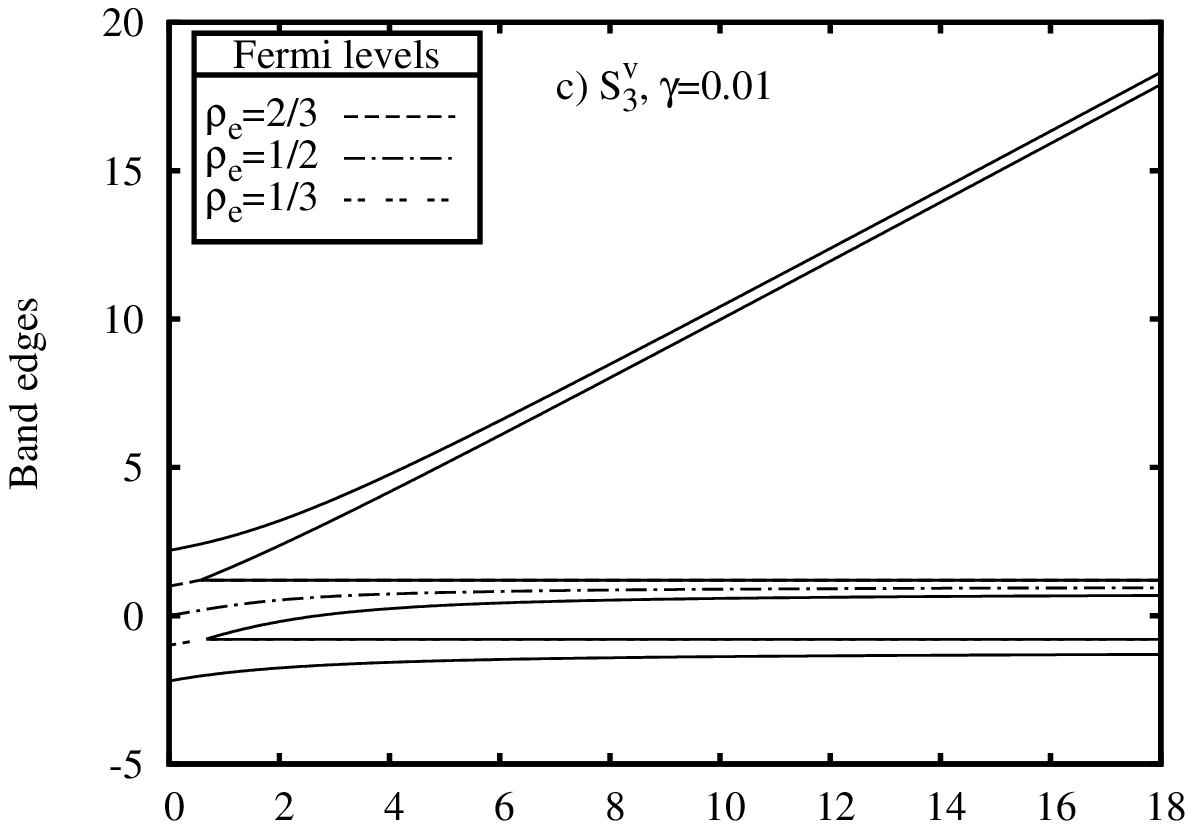}
\hfill
\includegraphics[width=0.41\textwidth]{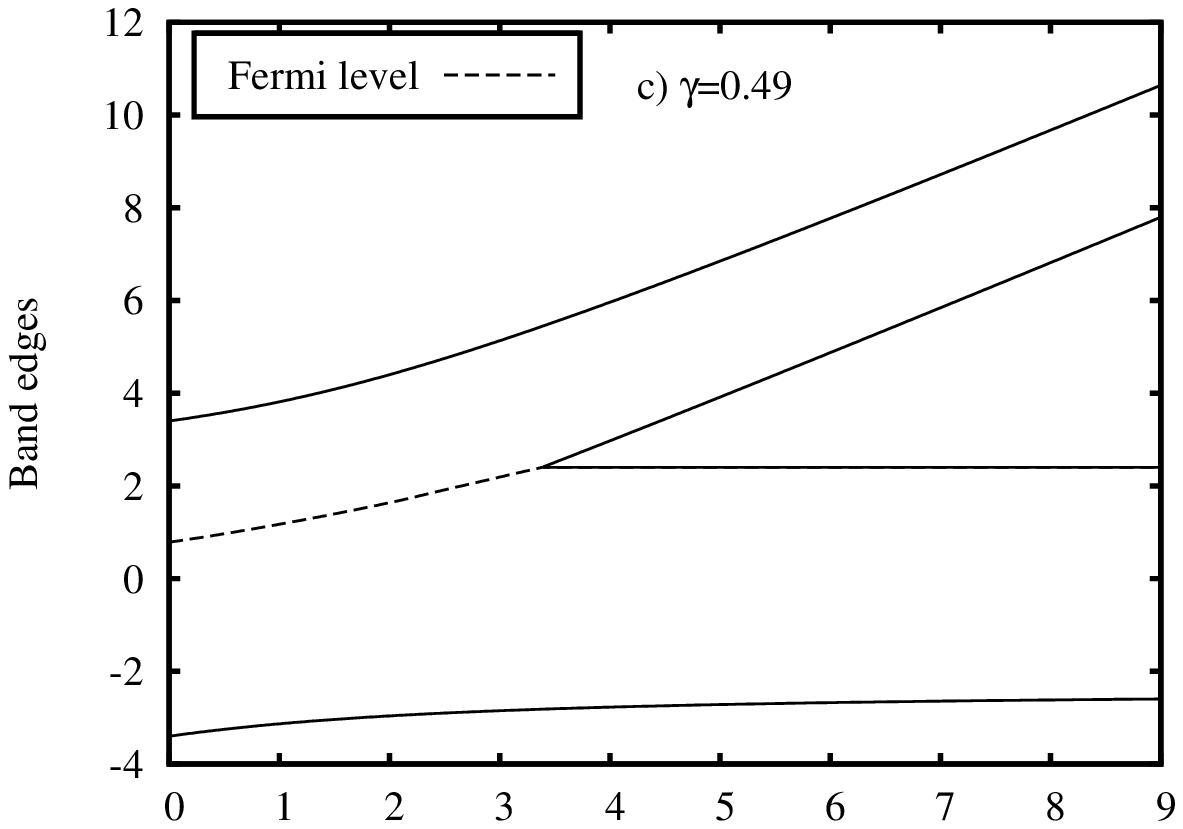}\\
\includegraphics[width=0.41\textwidth]{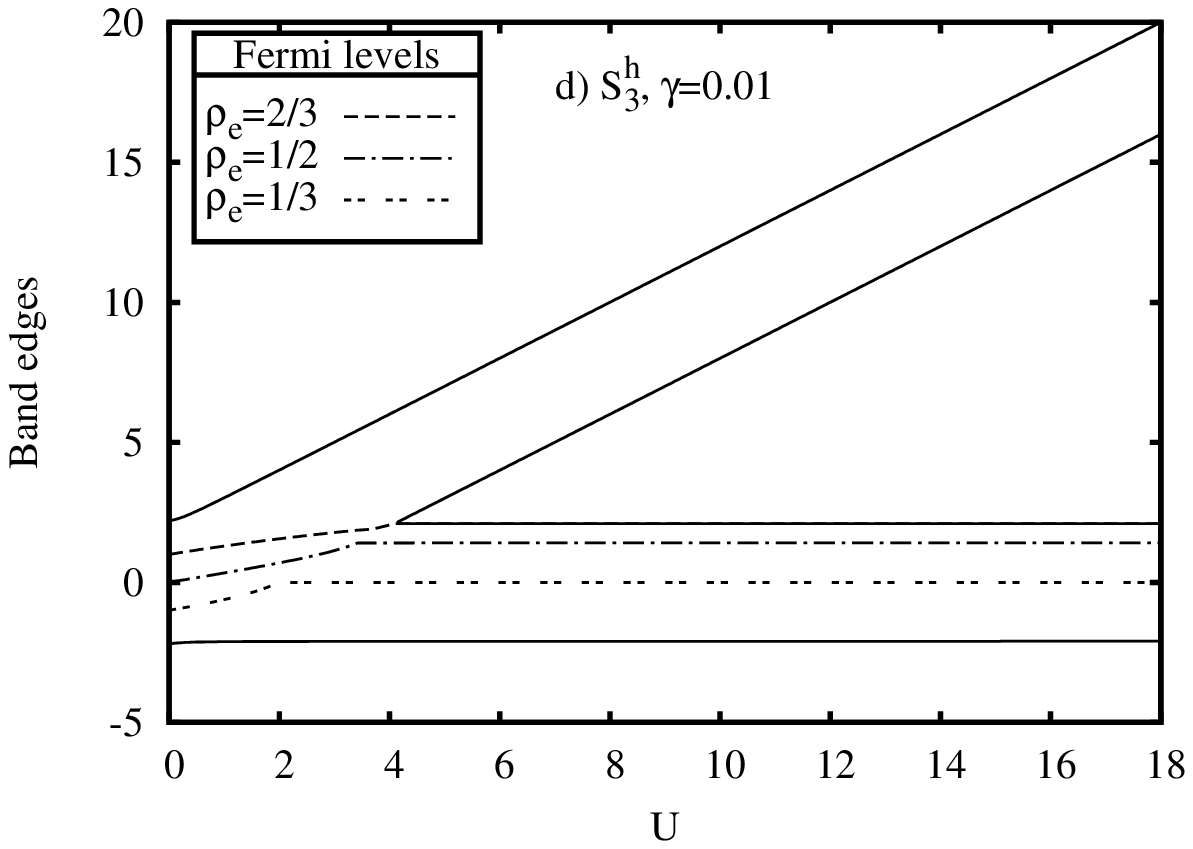}
\hfill
\includegraphics[width=0.41\textwidth]{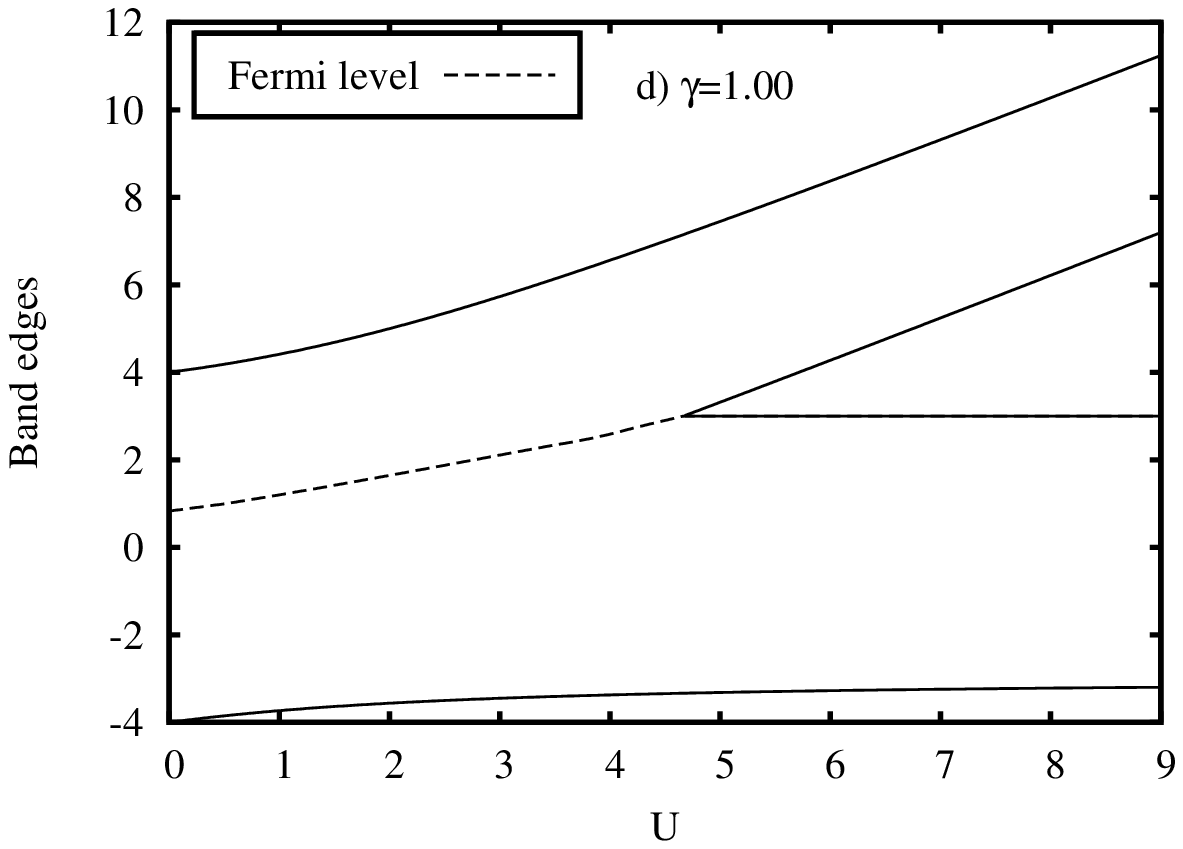}
\vspace{-5mm}
\\
\parbox[t]{0.47\textwidth}{\caption{{\small{Band edges and Fermi levels for configurations
$\mathcal{S}_{3}^{v}$ and $\mathcal{S}_{3}^{h}$.
Solid lines represent the band edges that are separated by a gap (note that in cases a), b)
and d) there is no gap between the first and the second band) while
dashed, dash-dotted and dotted lines -- the Fermi levels.
When a line of a Fermi level joins an intersection of two solid lines, then
it should be continued along the lower of the solid lines.
These plots correspond to the cases shown in Fig.~\ref{hfgse}
and Fig.~\ref{ohfgse}.}}}\label{bs34}}
\hfill
\parbox[t]{0.47\textwidth}{\caption{{\small{Band edges and Fermi levels for configuration $\mathcal{S}_{3}^{v}$
with electron density $\rho_e=2/3$,. Solid lines represent the band edges
(note that in cases c) and d) there is no gap between the first and the second band)
and the dashed line -- the Fermi level.
When a line of a Fermi level joins an intersection of two solid lines, then
it should be continued along the lower of the solid lines.
In the case a) the Fermi-level line coincides with the upper edge of the second band.
These plots correspond to the cases shown in Fig.~\ref{rele}.}}} \label{bs5}}
\end{figure}

It might be helpful to represent the above formulae graphically. We have
chosen to plot the band edges separated by a gap and the Fermi levels
as functions of $U$.
In Figs.~\ref{bs34} and ~\ref{bs5}, these plots are displayed for ion
configurations ${\mathcal{S}}^{v}_{3}$ and ${\mathcal{S}}^{h}_{3}$,
in all the cases considered in previous section.
Let $\{ |{\bf{k}}\rangle_{i} \}$ be a plane-wave basis of the space of single electron states,
labelled by wave-vectors ${\bf{k}}$ and index $i=1,\ldots,l$ enumerating the sublattices
of the underlying periodic ion configuration.
Let $h_{{\bf{k}}}$ be the matrix, in the plane-wave basis,
of the Hamiltonian of a single electron in an external field specified by
an axial-stripe configuration of horizontal period $l$ (which amounts to the number of sublattices).
The components of this matrix are labelled by the sublattices of the underlying periodic
ion configuration.
The matrix has the following structure:
\[
h_{{\bf{k}}}=
\begin{pmatrix}
\varepsilon_{v}+C_{1} & f_{1,2}(k_{h}) & \ldots & f_{1,l}(k_{h}) \\
f_{2,1}(k_{h}) & \varepsilon_{v}+C_{2} & \ldots & f_{2,l}(k_{h}) \\
\ldots & \ldots & \ddots & \ldots \\
f_{l,1}(k_{h}) & f_{l,2}(k_{h}) & \ldots &
\varepsilon_{v}+C_{l}
\end{pmatrix},
\]
where the matrix elements $f_{i,j}(k_h)$ are some functions of wave vector component $k_h$,
and $C_{i}$ are independent of the wave vector. In this matrix, the only elements which depend on wave vector
component $k_v$ are the diagonal elements, $(h_{{\bf{k}}})_{ii}=
\varepsilon_{v}+C_{i}$. Then, the solutions of the characteristic equation
assume the form:
$\lambda({\bf{k}})=\varepsilon_{v}+ f(k_h),$
with $f$ being some function of the horizontal component of ${\bf{k}}$.
Consequently, the equations for eigenvectors $\alpha$ with components $\alpha_i$,
$\sum_{i} (h_{{\bf{k}}}-\lambda({\bf{k}}))_{ij}\alpha_{j}=0,$
and so their solutions, do not depend on the vertical component of ${\bf{k}}$.

\end{document}